\documentclass{elsarticle}\usepackage[]{graphicx}\usepackage[]{color}
\makeatletter
\def\maxwidth{ %
  \ifdim\Gin@nat@width>\linewidth
    \linewidth
  \else
    \Gin@nat@width
  \fi
}
\makeatother

\definecolor{fgcolor}{rgb}{0.345, 0.345, 0.345}
\newcommand{\hlnum}[1]{\textcolor[rgb]{0.686,0.059,0.569}{#1}}%
\newcommand{\hlstr}[1]{\textcolor[rgb]{0.192,0.494,0.8}{#1}}%
\newcommand{\hlcom}[1]{\textcolor[rgb]{0.678,0.584,0.686}{\textit{#1}}}%
\newcommand{\hlopt}[1]{\textcolor[rgb]{0,0,0}{#1}}%
\newcommand{\hlstd}[1]{\textcolor[rgb]{0.345,0.345,0.345}{#1}}%
\newcommand{\hlkwb}[1]{\textcolor[rgb]{0.69,0.353,0.396}{#1}}%
\newcommand{\hlkwc}[1]{\textcolor[rgb]{0.333,0.667,0.333}{#1}}%
\newcommand{\hlkwd}[1]{\textcolor[rgb]{0.737,0.353,0.396}{\textbf{#1}}}%

\usepackage{framed}
\makeatletter
\newenvironment{kframe}{%
 \def\at@end@of@kframe{}%
 \ifinner\ifhmode%
  \def\at@end@of@kframe{\end{minipage}}%
  \begin{minipage}{\columnwidth}%
 \fi\fi%
 \def\FrameCommand##1{\hskip\@totalleftmargin \hskip-\fboxsep
 \colorbox{shadecolor}{##1}\hskip-\fboxsep
     \hskip-\linewidth \hskip-\@totalleftmargin \hskip\columnwidth}%
 \MakeFramed {\advance\hsize-\width
   \@totalleftmargin\z@ \linewidth\hsize
   \@setminipage}}%
 {\par\unskip\endMakeFramed%
 \at@end@of@kframe}
\makeatother

\definecolor{shadecolor}{rgb}{.97, .97, .97}
\definecolor{messagecolor}{rgb}{0, 0, 0}
\definecolor{warningcolor}{rgb}{1, 0, 1}
\definecolor{errorcolor}{rgb}{1, 0, 0}
\newenvironment{knitrout}{}{} 

\usepackage{alltt}         

\usepackage{lineno}
\modulolinenumbers[5]

\journal{NeuroImage}

\usepackage{cprotect}
\usepackage{makecell}

\usepackage{booktabs} 
\usepackage{amsmath,amsfonts,amssymb}   

\usepackage{natbib}
\biboptions{authoryear}

\setcitestyle{round}

\makeatletter
\def\ps@pprintTitle{%
   \let\@oddhead\@empty
   \let\@evenhead\@empty
   \let\@oddfoot\@empty
   \let\@evenfoot\@oddfoot
}
\makeatother
\IfFileExists{upquote.sty}{\usepackage{upquote}}{}
\begin{document}

\title{Meta-Analysis of Generalized Additive Models in Neuroimaging Studies}

\begin{frontmatter}

\author[1]{{\O}ystein S{\o}rensen\corref{corrauth}}
\cortext[corrauth]{Corresponding author: {\O}ystein S{\o}rensen, Department of Psychology, Pb. 1094 Blindern, 0317 Oslo, Norway.}
\ead{oystein.sorensen@psykologi.uio.no}

\author[4,5]{Andreas M Brandmaier}
\author[3]{D{\'{i}}dac Maci{\`{a}}}
\author[6]{Klaus Ebmeier}
\author[7,8,9]{Paolo Ghisletta}
\author[10]{Rogier A Kievit}
\author[1]{Athanasia M Mowinckel}
\author[1,2]{Kristine B Walhovd}
\author[1]{Rene Westerhausen}
\author[1,2]{Anders Fjell}

\address[1]{Center for Lifespan Changes in Brain and Cognition, University of Oslo, Norway}
\address[2]{Department of Radiology and Nuclear Medicine, Oslo University Hospital, Norway}
\address[3]{Departament de Medicina, Facultat de Medicina i Ci{\`{e}}ncies de la Salut, Universitat de Barcelona, and Institut de Neuroci{\`{e}}ncies, Universitat de Barcelona, Spain}
\address[4]{Center for Lifespan Psychology, Max Planck Institute for Human Development, Berlin, Germany}
\address[5]{Max Planck UCL Centre for Computational Psychiatry and Ageing Research, Berlin, Germany}
\address[6]{Department of Psychiatry, University of Oxford, UK}
\address[7]{Faculty of Psychology and Educational Sciences, University of Geneva, Switzerland}
\address[8]{Swiss Distance University Institute, Switzerland}
\address[9]{Swiss National Centre of Competence in Research LIVES, University of Geneva, Switzerland}
\address[10]{MRC Cognition and Brain Sciences Unit, University of Cambridge, UK}

\begin{abstract}
Analyzing data from multiple neuroimaging studies has great potential in terms of increasing statistical power, enabling detection of effects of smaller magnitude than would be possible when analyzing each study separately and also allowing to systematically investigate between-study differences. Restrictions due to privacy or proprietary data as well as more practical concerns can make it hard to share neuroimaging datasets, such that analyzing all data in a common location might be impractical or impossible. Meta-analytic methods provide a way to overcome this issue, by combining aggregated quantities like model parameters or risk ratios. Most meta-analytic tools focus on parametric statistical models, and methods for meta-analyzing semi-parametric models like generalized additive models have not been well developed. Parametric models are often not appropriate in neuroimaging, where for instance age-brain relationships may take forms that are difficult to accurately describe using such models. In this paper we introduce meta-GAM, a method for meta-analysis of generalized additive models which does not require individual participant data, and hence is suitable for increasing statistical power while upholding privacy and other regulatory concerns. We extend previous works by enabling the analysis of multiple model terms as well as multivariate smooth functions. In addition, we show how meta-analytic $p$-values can be computed for smooth terms. The proposed methods are shown to perform well in simulation experiments, and are demonstrated in a real data analysis on hippocampal volume and self-reported sleep quality data from the Lifebrain consortium. We argue that application of meta-GAM is especially beneficial in lifespan neuroscience and imaging genetics. The methods are implemented in an accompanying R package \verb!metagam!, which is also demonstrated.

\section*{Highlights}
\begin{itemize}
\item allows combination of nonlinear models without sharing data.

\item increases power and accuracy in neuroimaging studies.

\item illustrated in case study from the Lifebrain consortium.

\item is available in open source R package.

\end{itemize}

\end{abstract}

\begin{keyword}
data protection \sep distributed learning \sep generalized additive mixed models \sep generalized additive models \sep meta-analysis \sep privacy
\end{keyword}

\end{frontmatter}

\section{Introduction}

Combining brain imaging data across studies has great potential in terms of increasing statistical power, enabling discoveries of effects that might not be detectable in any single dataset. Due to regulatory and practical concerns, privacy in particular, it may not be possible to analyze all data in a single place. It may also sometimes be beneficial to analyze data from multiple studies in two stages, even when the data are available at a single location, e.g., when data do not fit in computer memory or runtime is nonlinear in the number of participants \citep{Riley2010}.

Meta-analytic techniques offer one way to increase statistical power without sharing raw data. By estimating the relationships under study separately in each data location, pooled estimates are obtained by combining the estimates without sharing the underlying data. With some exceptions, meta-analytic methods have been developed for combining parameters from parametric statistical models or for effect measures like relative risks \citep{Hedges1985,Sutton2008}. However, there are important cases in which it is impractical and suboptimal to enforce a parametric representation of the association under investigation, e.g., when an appropriate parametric model to approximate the data is not known, or its interpretability is not clear, as with high-degree polynomials. Examples include lifetime trajectories of brain development \citep{Fjell2010}, air quality measures \citep{Gasparrini2010}, and ecological phenomena \citep{Borchers1997,Pedersen2019}. Generalized additive models (GAMs) \citep{Hastie1986,Wood2017} are attractive for studying such relationships, and can easily be extended to longitudinal or other forms of clustered data via generalized additive mixed models (GAMMs), which, in addition to GAMs, can also estimate random effects.

\begin{figure}
\centering
\includegraphics[width=.8\linewidth]{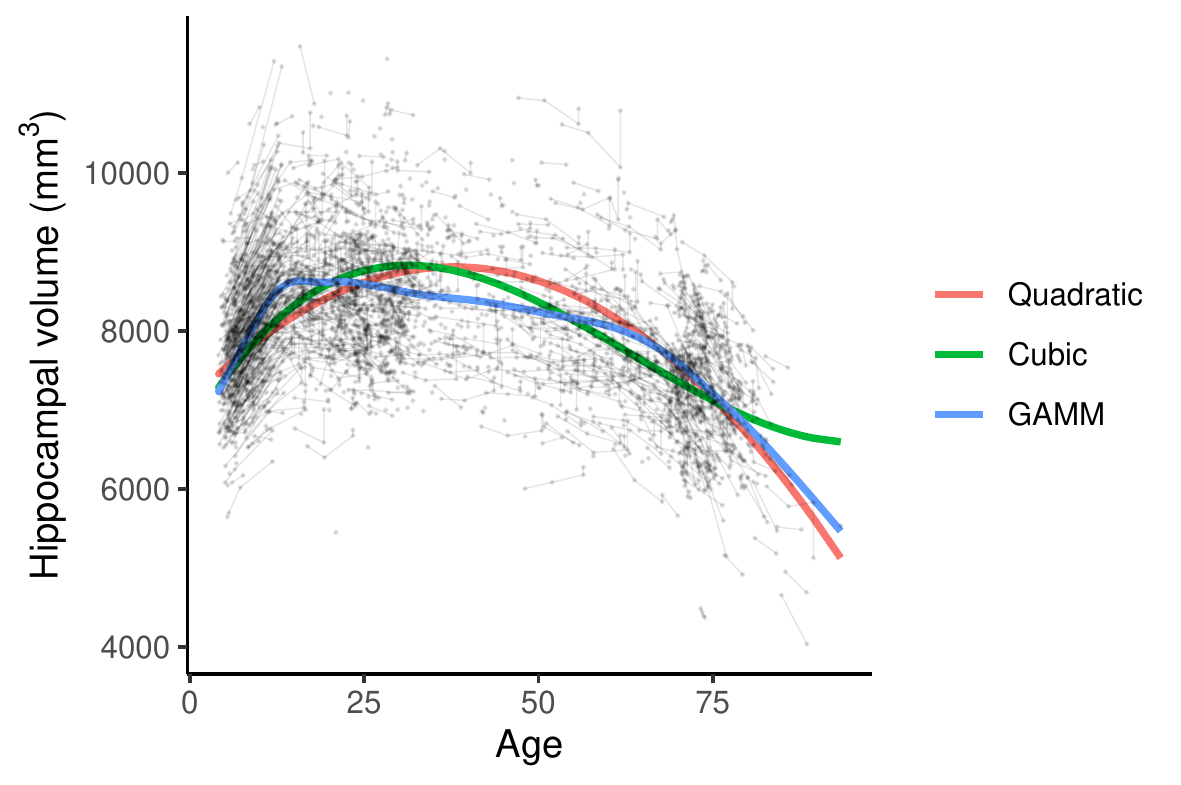}
\cprotect\caption{\textbf{Modeling lifespan trajectories.} Example of modeling lifespan hippocampal volume with longitudinal data using linear mixed models with quadratic and cubic terms for age, as well as a generalized additive model. The black dots show individual observations and the black lines connect subsequent observations from the same individual. The GAMM was fitted with 20 cubic regression splines and a random intercept term for each individual, and the optimal smoothing parameter estimated with restricted maximum likelihood.}
\label{fig:gamm_vs_poly}
\end{figure}

Figure \ref{fig:gamm_vs_poly} illustrates modeling lifespan trajectories of hippocampal volume changes using linear mixed models (LMMs) with quadratic and cubic polynomials for the age term, and a GAMM with a smooth term for age\cprotect\footnote{The LMMs were fitted using R \citep{Rcore} package \verb!nlme! \citep{Pinheiro2019} and the GAMM was fitted using \verb!mgcv! \citep{Wood2017}, all with a random intercept term.}. The data were taken from 4,364 observations of 2,023 healthy participants (age 4-93 years, 1-8 measurements per participant) from the Center for Lifespan Changes in Brain and Cognition (LCBC) longitudinal studies \citep{Walhovd2016,Fjell2017}. Detailed sample characteristics are presented in Supplementary Material I. The quadratic fit is not flexible enough to capture the steep increase during adolescence - moreover, it estimates the hippocampal volume to increase until the age of around 40. The cubic fit captures the volume growth during adolescence better than the quadratic fit, but fails to capture the decline that occurs after the age of around 70. The GAMM fit, on the other hand, is flexible enough to both capture the steep increase during adolescence, a period of moderate decline during adulthood, and finally a steeper decline at older age\cprotect\footnote{Figure \ref{fig:gamm_vs_poly} and all other figures in this paper were created using \verb!ggplot2! \citep{Wickham2016}.}.

As the methods for meta-analysis of GAMs and GAMMs are identical, we will refer to both as GAMs in the rest of this paper, unless distinction is necessary. For reasons that we will explain below, in this paper we will not discuss meta-analysis of the underlying parametric functions across GAMs. Rather, we present methods for combining GAM fits for neuroimaging data by pointwise meta-analysis of the fitted values. Although developed for use in meta-analytic neuroimaging studies, the methods can of course be applied to other types of data as well. The models under study can include any number of terms, including multivariate smooth functions. In order to employ these techniques, models should be fit separately for each cohort, with basis functions and knot placement chosen independently. Related previous works include meta-analysis of locally weighted regression fits \citep{Schwarz2000} and meta-analytic estimation of nonlinear dose-response relationships using individual participant data \citep{Crippa2018,Sauerbrei2011}.

The main applications we have in mind are meta-analysis of published results where the effects of interest are represented by functional relationships rather than single parameters, and multi-center studies in which it is impractical or not possible to analyze all brain imaging data in a single location. An example of the latter is the Enhancing Neuro Imaging Genetics through Meta Analysis project (ENIGMA: http://enigma.ini.usc.edu/), where meta-analysis of individual site summary statistics is the commonly applied strategy (e.g., \citet{Dennis2018,vanErp2018}). The methods developed require that some model relating an outcome of interest to a set of explanatory variables has been fitted on data from each cohort, and that the model estimates can be shared across cohorts such that the expected response and their standard errors at new values of the explanatory variables can be computed. We provide a companion R package named \verb!metagam! \citep{Sorensen2020} containing functions for removing all individual participant data from GAMs fitted with the \verb!mgcv! and \verb!gamm4! packages \citep{Wood2017,Wood2017gamm4}, such that the resulting model object only contains aggregate measures which can easily be shared. The package also contains methods for combining the fits and analyzing the results, and will be demonstrated in Section \ref{sec:metagam}. The comprehensive review of meta-analysis packages in R by \citet{Polanin2017} does not mention any existing packages for conducting this type of pointwise meta-analysis, so to the best of our knowledge, \verb!metagam! is the first R package to provide this functionality.

The methods presented in this paper were motivated by a project in the Lifebrain consortium (http://www.lifebrain.uio.no/) \citep{Walhovd2018}. The goal was to study the relationship between self-reported sleep and hippocampal volume across six Lifebrain cohorts, and GAMMs were a natural model choice due to the expected non-linear age-relationships for self-reported sleep parameters and hippocampal volume. In this case a safe common data store was in place, but we initially hypothesized that it might be easier to have each cohort fit a model locally and share the overall result rather than analyzing all data in a single place, leading to the development of the methods presented here.

\section{Background}

\subsection{Meta-Analysis of Parametric Models}
Consider a situation in which $M$ cohorts $m=1,\dots,M$ each have a dataset $D_{m}$ with $n_{m}$ participants. The response variable of interest is denoted $y$ and there are $p$ explanatory variables represented by the vector $\mathbf{x}$. If subject $i$ in cohort $m$ has been measured $n_{mi}$ times, the data are $D_{m} = \{(y_{ij}, \mathbf{x}_{ij})$, \text{ for } $i=1,\dots,n_{m}$, $j=1,\dots,n_{mi}\}$. Notably, this includes the case of individually varying numbers of assessments and time intervals between assessments. In practice, some of the explanatory variables will be time-varying, while others will be time-invariant. Purely cross-sectional data correspond to $n_{mi}=1$ for all $m$ and $i$.

Our interest concerns statistical inference on data from all studies, in the case where data cannot be analyzed jointly. When the relationship under study can be represented by a parametric model, well established methods exist for obtaining meta-analytic estimates of the model parameters. For example, if an LMM is used for longitudinal data \citep{Laird1982}, parameter estimates from each study can be combined using parametric meta-analysis \citep{derSimonian1986,Gasparrini2012}. The same applies to related approaches based on structural equation modeling (e.g., \citet{Brandmaier2018,Kievit2018}) or generalized linear models \citep{McCullagh1989}.

\subsection{Generalized Additive Models}
\label{sec:gam}

In many applications, assuming that the response%
\cprotect\footnote{For ease of presentation, we assume a continuous outcome with normally distributed residuals, corresponding to an identity link function in a generalized additive model. The methods developed extend directly to other outcomes (e.g., binomial or count) by introducing a linear predictor $\eta = \beta_{0} + \sum_{s=1}^{S} f_{s}(\mathcal{X}_{s})$ with link function $g(\cdot)$ satisfying $g(y) = \eta$.}
$y$ is a smooth function of the explanatory variables, rather than following a model that is linear in its parameters (e.g., polynomial), may lead to better statistical fit, cf. Figure \ref{fig:gamm_vs_poly}. Generalized additive models (GAMs) \citep{Hastie1986} take this approach. Letting $\mathcal{X}_{s}$ denote the set of explanatory variables used by smooth function $f_{s}(\cdot)$, a GAM with $S$ smooth terms can be written on the form
\begin{equation}
\label{eq:gam_generic}
y = \beta_{0} + \sum_{s=1}^{S} f_{s}\left(\mathcal{X}_{s}\right) + \epsilon,
\end{equation}
where $\beta_{0}$ denotes the intercept and $\epsilon$ is a normally distributed residual. Constraints necessary for model identification are discussed in \ref{sec:smooth_constraints}. Each smooth function is a linear combination of $K_{s}$ basis functions $b_{ks}(\cdot)$ with weights $\gamma_{ks}$, $k=1,\dots,K_{s}$,
\begin{equation}
\label{eq:gam_weights}
f_{s}\left(\mathcal{X}_{s}\right) = \sum_{k=1}^{K_{s}} b_{ks}(\mathcal{X}_{s}) \gamma_{ks}.
\end{equation}
Typically, each basis function is nonzero over a small part of the range of its variables, as defined by its knot locations. A linear parametric term for $x_{j}$ is given by the special case $\mathcal{X}_{s} = \{x_{j}\}$, $K_{s}=1$, $b_{1s}(x_{j})=x_{j}$, and  hence $f_{s}(\mathcal{X}_{s}) = \gamma_{1s} x_{j}$. Examples are provided in Supplementary Material II.

\subsection{Smoothing} 

Least squares estimation of model \eqref{eq:gam_generic} with a large number of basis functions for each term typically leads to wiggly estimates which overfit the data. Smoothing is thus necessary, and a popular and efficient solution involves penalizing the second derivatives of the smooth functions, while making sure the number of basis functions is sufficiently large to represent a wide range of functional forms \citep{Wood2017}. In the context of meta-analysis, smoothing is performed independently for each study. Supplementary Material II presents further details and a visualization of the effect of smoothing.

\subsection{Limitations of Parametric Meta-Analysis of Generalized Additive Models}
\label{sec:knot_placement}

If each study used identical basis functions, a meta-analytic fit could be obtained by treating their weights as linear regression parameters \citep{Gasparrini2012}. However, as also noted by \citet{Crippa2018}, if the range of some variable $x_{j}$ differs between cohorts, enforcing the same knot placement is suboptimal and the model may not even be identified.

As an example, we consider modeling of lifespan trajectories of hippocampal volumes from six European cohorts. The data are further described in Section \ref{sec:case_study}. As shown in Figure \ref{fig:cohort_dist} (top), these studies have widely varying age distributions. We fit GAMs relating baseline age to hippocampal volume for each cohort, but enforced the same knot location for all models, placed at eight equally spaced quantiles of the full data sample. Table \ref{tab:equal_knots_meta} shows the corresponding spline coefficients. While these coefficients are not directly interpretable, outliers for a given sample indicate that its fit is highly different from the others, and a missing value indicates that the fit for the sample was not identified. As can be seen, Barcelona and Whitehall-II have missing values (-) for spline coefficient $\gamma_{2}$. In addition, there are extreme outliers: Barcelona has a severely outlying value for $\gamma_{1}$, BASE-II has an outlying value for $\gamma_{8}$, and Whitehall-II has outlying values for $\gamma_{1}$ and $\gamma_{3}$. This lack of identification and unstable coefficients is caused by using knot locations which, because they are forced to be equal across cohorts, are not suitable for the actual age distributions.

\begin{table}
\centering
\small
\begin{tabular}{lrrrrrrrr}
  \hline
Study & $\gamma_{1}$ & $\gamma_{2}$ & $\gamma_{3}$ & $\gamma_{4}$ & $\gamma_{5}$ & $\gamma_{6}$ & $\gamma_{7}$ & $\gamma_{8}$ \\ 
  \hline
Barcelona & 28142 & - & 6195 & 7719 & 7629 & 7421 & 7190 & 6310 \\ 
  BASE-II & 4694 & 8374 & 6274 & 7919 & 7770 & 7213 & 7297 & -17182 \\ 
  Betula & 9605 & 8481 & 8380 & 8072 & 7840 & 7389 & 6994 & 7734 \\ 
  Cam-CAN & 8298 & 8452 & 8397 & 8040 & 7916 & 7468 & 7291 & 6375 \\ 
  LCBC & 8408 & 8479 & 8324 & 7689 & 7401 & 7468 & 7202 & 5819 \\ 
  Whitehall-II & 1625151 & - & -120033 & 7580 & 7528 & 7353 & 6935 & 6084 \\ 
   \hline
\end{tabular}
\cprotect\caption{Spline coefficients for models described in Section \ref{sec:knot_placement}. The coefficient $\gamma_{2}$ was not possible to determine for Barcelona and Whitehall-II. In addition, $\gamma_{1}$ for Barcelona and Whitehall-II, $\gamma_{3}$ for Whitehall-II, and $\gamma_{8}$ for BASE-II are severe outliers.}
\label{tab:equal_knots_meta}
\end{table}

\section{Pointwise Meta-Analysis of Generalized Additive Models}
\label{sec:methods}

\subsection{Estimation of Overall Fits in Pointwise Meta-Analysis}

We now propose a model for meta-analysis of GAMs. We assume that a GAM has been fitted to the data from each cohort $m$ separately, and that the vector $\mathbf{x}$ represents values of the explanatory variables for which a meta-analytic estimate of the regression function is sought. The expected response in cohort $m$ is then given by
\begin{equation}
\label{eq:fitted_model}
\hat{y}_{m} = \hat{f}_{m}\left(\mathbf{x}\right) = \hat{\beta}_{0, m} + \sum_{s=1}^{S}\hat{f}_{s,m}\left(\mathcal{X}_{s}\right), ~ m=1,\dots,M.
\end{equation}
Importantly, the basis functions and knot placements for a given smooth term $\hat{f}_{s,m}\left(\mathcal{X}_{s}\right)$ will in general vary across cohorts $m$. Each model term has a corresponding estimated standard deviation $\hat{\sigma}_{s,m}(\mathcal{X}_{s})$, and the overall fit has estimated standard deviation $\hat{\sigma}_{m}(\mathbf{x})$.

We illustrate our methods by considering meta-analytic estimation of each single term separately, but note that inference on any combination of smooth terms, including the overall function, is readily obtained with the same methods. Some additional details related to identification of smooth terms are discussed in \ref{sec:smooth_constraints}. For ease of notation, we omit the dependency on $\mathcal{X}_{s}$ and $\mathbf{x}$ in the rest of this section. For example, $f_{s,m}$ means $f_{s,m}(\mathcal{X}_{s})$ and $\sigma_{s,m}$ means $\sigma_{s,m}(\mathcal{X}_{s})$. 

The meta-analytic estimate of smooth term $s$ is the weighted mean
\begin{equation}
\label{eq:meta_est}
\hat{f}_{s} = \frac{\sum_{m=1}^{M} \hat{f}_{s,m} \left(\hat{\sigma}_{s,m}^{2} + \hat{\sigma}_{s}^{2}\right)^{-1}}{\sum_{m=1}^{M} \left(\hat{\sigma}_{s,m}^{2} + \hat{\sigma}_{s}^{2} \right)^{-1}}
\end{equation}
with standard error
\begin{equation}
\label{eq:meta_se}
\text{se}_{\hat{f}_{s}} = \left\{\sum_{m=1}^{M} \hat{\sigma}_{s,m}^{2} + \hat{\sigma}_{s}^{2}\right\}^{-1/2}.
\end{equation}
The term $\hat{\sigma}_{s}^{2}$ represents the estimated between-study variance, and fixed effects meta-analysis corresponds to the special case $\hat{\sigma}_{s}^{2} = 0$. The DerSimonian-Laird estimator for between-sample variance \citep{derSimonian1986},
\begin{equation}
\label{eq:between_var_estimator}
\hat{\sigma}_{s}^{2} = \text{max}\left\{ 0, \frac{\sum_{m=1}^{M}\hat{\sigma}_{s,m}^{-2}\left(\hat{f}_{s,m} - \frac{\sum_{m=1}^{M} \hat{\sigma}_{s,m}^{-2}\hat{f}_{s,m} }{ \sum_{m=1}^{M}\hat{\sigma}_{s,m}^{-2} } \right) - \left(M - 1\right)}{\sum_{m=1}^{M} \hat{\sigma}_{s,m}^{-2} - \frac{\sum_{m=1}^{M} \hat{\sigma}_{s,m}^{-4} }{ \sum_{m=1}^{M} \hat{\sigma}_{s,m}^{-2}} }\right\},
\end{equation}
is computationally efficient as it does not require iteration, making it attractive in pointwise meta-analysis in which a separate estimate is required over a large number of grid points. However, iterative methods may give more accurate estimates \citep{Veroniki2016}. We refer to, e.g. \citet{Viechtbauer2005} and \citet{Viechtbauer2015} for an overview of estimators of between-sample variance, all of which can be used with the methods presented.

Equations \eqref{eq:meta_est}-\eqref{eq:meta_se} are the familiar weighted means formulas used in meta-analysis, and have been used by \citet{Sauerbrei2011} in a similar setting, focusing on meta-analysis of univariate functions estimated by fractional polynomials. In the fixed effects case, $\hat{f}_{s}$ is the estimated mean conditional on randomly pooling from the populations of the observed cohorts alone. Random effects analysis, on the other hand, estimates the marginal population effect $f_{s}$ across all potential studies. See \citet[Sec. 2.3]{Viechtbauer2010} for an excellent discussion of the interpretation of fixed vs. random effects meta-analyses. Confidence bands with level $(1 - \alpha)$ are readily obtained for either estimates as
\begin{equation}
\label{eq:confint}
\left[\hat{f}_{s} + z_{\alpha/2} \text{se}_{\hat{f}_{s}}, \hat{f}_{s} + z_{1-\alpha/2} \text{se}_{\hat{f}_{s}}\right],
\end{equation}
where $z_{q}$ denotes the $q$th quantile of the standard normal distribution. 

Pointwise meta-analysis requires software for computing predictions and standard errors for the models fitted in each study. In the case of GAMs, this requires knowledge of the basis functions along with the estimates and covariance matrices of spline weights, quantities which are readily available from software for fitting GAMs, like \verb!mgcv! \citep{Wood2017} or \verb!pyGAM! \citep{Serven2018}. Importantly, individual participant data are not required for computing such predictions from already fitted models.

\subsection{Inference for Smoothing Terms in Pointwise Meta-Analysis}

Tests for statistical significance of smooth terms can be performed by combining the $p$-values from each separate fit using methods for meta-analytic combination of $p$-values as summarized, e.g., in \citet{Becker1994} or \citet{Loughin2004}. In particular, let $p_{s,m}$ denote the $p$-value obtained in cohort $m$ for the hypothesis $H_{0,m}:f_{s}(\mathcal{X}_{s}) = 0$ that the smooth term $s$ is zero over the whole range of explanatory variables $\mathcal{X}_{s}$ in cohort $m$, and let $H_{A,m}:f_{s}(\mathcal{X}_{s})\neq 0$ denote the alternative hypothesis. Such $p$-values can be computed using the methods in \citet{Wood2012}. The meta-analytic null hypothesis then states that all $p$-values are uniformly distributed between 0 and 1, i.e., $H_{0}: p_{s,m} \sim U(0, 1)$, $m=1,\dots,M$, while the meta-analytic alternative hypothesis $H_{A}$ states that all $p$-values have the same unknown non-uniform density which is non-increasing in the test statistic \citep{Birnbaum1954}. A large number of methods exist for computing the combined $p$-values. For example, Stouffer's sum of z method \citep{Stouffer1949} uses the Z-score
\begin{equation}
\label{eq:Stouffer}
Z_{s} = \frac{\sum_{m=1}^{M} w_{m}\Phi^{-1}\left(1-p_{s,m}\right)}{\sqrt{\sum_{m=1}^{M} w_{m}^{2}}},
\end{equation}
where $\Phi$ is the standard normal distribution and $\Phi^{-1}$ its quantile function, and $w_{m}, m=1,\dots,M$ are meta-analytic weights. \citet{Zaykin2011} suggests defining the weights as the square root of the sample size, $w_{m} = \sqrt{n_{m}}$. The combined $p$-value is then defined by $p_{s} = 1 - \Phi(Z_{s})$.

\section{Simulation Studies}
\label{sec:simsec}

Simulation studies were conducted in order to compare the performance of the pointwise meta-analysis approach presented in Section \ref{sec:methods} to the ideal mega-analysis \citep{McArdle1985} case, in which all data can be analyzed jointly. Section \ref{sec:simulation} reports simulation results comparing estimation of smooth terms, and Section \ref{sec:power} reports simulation results comparing statistical inference performance.

\subsection{Function Estimation}
\label{sec:simulation}

\begin{figure}
\centering
\includegraphics[width=\columnwidth]{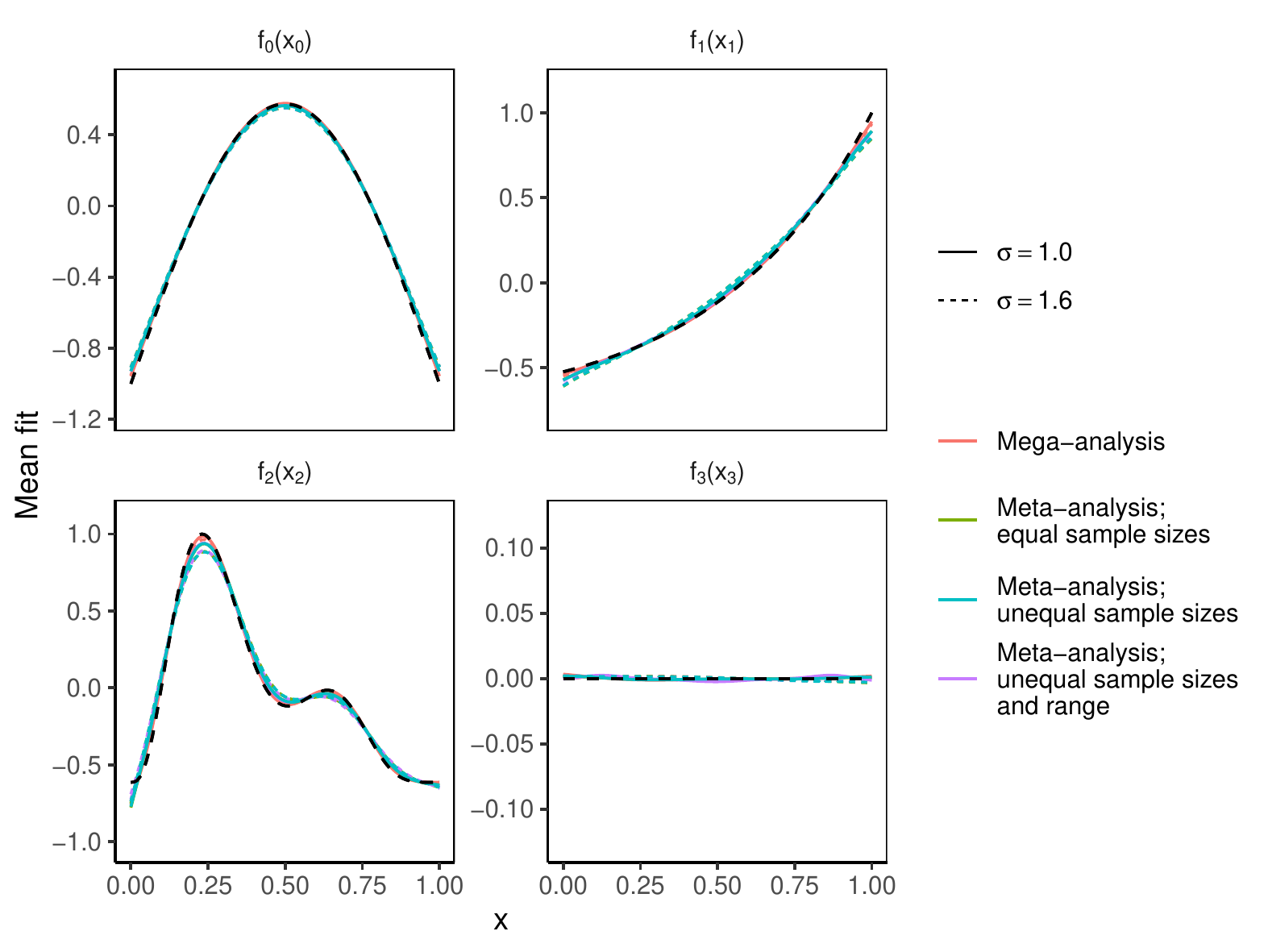}
\caption{\textbf{Simulation estimates overlaid on true functions.} Dashed black lines show true functions. The colored lines show mean fits averaged over 1,000 simulations as described in Section \ref{sec:simulation}.}
\label{fig:univariate_fits}
\end{figure}

The first set of simulations compared pointwise meta-analysis to mega-analysis in terms of their ability to accurately estimate nonlinear functional forms and to quantify uncertainty with confidence bands. Data were generated from the model
\begin{equation*}
y = f_{0}(x_{0}) + f_{1}(x_{1}) + f_{2}(x_{2}) + f_{3}(x_{3}) + \epsilon,
\end{equation*}
with all explanatory variables independently uniformly distributed in $[0, 1]$ and $\epsilon \sim N(0, \sigma^{2})$. The functional forms assumed were similar to those used by \citet{Marra2012}, and are shown as dashed black lines in Figure \ref{fig:univariate_fits}.

Datasets with 4,000 observations of $(x_{0}, x_{1}, x_{2}, x_{3}, y)$ were independently sampled 1,000 times. For each dataset, the following four cases were considered:

\begin{itemize}
\item In the mega-analysis case, all 4,000 observations were analyzed jointly. This served as a gold standard, yielding the model that would be fit if all data were available to analyze with a single model.
\item In the equal sample size case, the dataset was split into 5 "cohorts" of 800 observations each. Each cohort was analyzed independently, and the meta-analytic fit computed as outlined above.
\item In the unequal sample size case, the dataset was split into 5 "cohorts" with 300, 500, 800, 1,000, and 1,400 observations each.
\item In the unequal range and sample size case, a first "cohort" was created by sampling 300 observations with $x_{2} < 0.5$ from the full dataset, the second cohort by sampling 500 observations with $x_{2} \geq 0.5$ from the remaining observations, the third cohort by sampling 800 observations with $x_{1} < 0.5$ from the remaining observations, the fourth cohort by sampling 1,000 observations with $x_{1} \geq 0.5$ from the remaining observations, and the fifth cohort contained the remaining 1,400 observations. Hence, this case has the same sample sizes as the unequal sample size case, but the ranges of $x_{1}$ and $x_{2}$ vary between cohorts.
\end{itemize}

In the latter three cases, fixed effects meta-analysis was conducted. Univariate smooth terms were estimated using cubic regression splines with 20, 10, 30, and 5 basis functions for $f_{0}(x_{0})$, $f_{1}(x_{1})$, $f_{2}(x_{2})$, and $f_{3}(x_{3})$, respectively. Knot placement was determined independently for each cohort, based on the quantiles of the explanatory variables. Second derivative smoothing was performed using generalized cross-validation, and standard error computations for each term included the uncertainty about the overall intercept as described in \citet{Marra2012}. For identifiability, the smooth terms were subject to sum-to-zero constraints over $[0,1]$, cf. \ref{sec:smooth_constraints}. In the case study reported in Section \ref{sec:case_study}, with a GAM regressing hippocampal volume on age and sleep quality, the mega-analysis case had an adjusted R squared value ${R}_{adj}^2 = 0.37$ (cf. Supplementary Material IV, p. 13). Setting $\sigma = 1.0$ in the simulations gave $R_{adj}^2 \approx 0.40$, thus close to a realistic noise level in neuroimaging studies, while $\sigma = 1.6$ corresponds to a high noise case with $R_{adj}^2 \approx 0.20$. All simulations were repeated with each of these noise levels. Computations were performed in R version 3.6.2 \citep{Rcore} with the package \verb!mgcv! \citep{Wood2017}.

Figure \ref{fig:univariate_fits} shows the average fits over all simulations. One can hypothesize that splitting a dataset into smaller parts and performing smoothing separately might lead to oversmoothing compared to analyzing all data in a single model. Considering Figure \ref{fig:univariate_fits} we see that this was the case for estimating $f_{2}(x_{2})$ in the case with $\sigma=1.0$, in which all meta-analysis cases slightly underestimated the two peaks of the true term. For the three other terms, the $\sigma=1.0$ case had very low bias. In the high noise case, with $\sigma = 1.6$, oversmoothing can also be seen in the estimates of $f_{1}(x_{1})$. The two meta-analyses with unequal sample size, also had somewhat too smooth estimates of $f_{1}(x_{1})$ in the $\sigma=1.0$ case. Overall, however, the average fits in the meta-analysis cases were very close to the true curves.

Table \ref{tab:univariate_rmse} shows the root-mean-square error (RMSE) of the fitted terms over the range $[0, 1]$. In both noise settings, the meta-analyses with equal and unequal sample size had only slightly higher RMSE than the mega-analytic estimates, and there did not seem to be any systematic difference between them. The meta-analysis with unequal range and sample size had RMSE very close to the two other meta-analytic cases.

Table \ref{tab:univariate_coverage} shows the average coverage across $[0, 1]$ of 95 \% confidence intervals computed with \eqref{eq:confint}. The coverage of the confidence intervals of the mega-analytic estimates were close to 95 \%, as expected from \citet{Marra2012}, and always conservative. All three meta-analytic cases had very similar coverage, varying between 86 \% and 99 \%. In particular for $f_{1}(x_{1})$ and $f_{2}(x_{2})$ the confidence intervals were somewhat too narrow, whereas for $f_{0}(x_{0})$ and $f_{3}(x_{3})$ the confidence intervals were slightly conservative.

\begin{table}
\small
\centering
\begin{tabular}{ll|cccc}
\toprule
Term & $\sigma$ & \thead{Equal sample \\ size} & \thead{Unequal sample \\ size} & \thead{Unequal range \\ and sample size} & \thead{Mega-analysis} \\
\hline
$f_{0}(x_0)$ & 1.00 & 0.037 (0.011) & 0.037 (0.011) & 0.038 (0.012) & 0.035 (0.013) \\ 
  $f_{1}(x_1)$ & 1.00 & 0.037 (0.014) & 0.037 (0.014) & 0.040 (0.013) & 0.031 (0.011) \\ 
  $f_{2}(x_2)$ & 1.00 & 0.060 (0.011) & 0.060 (0.011) & 0.062 (0.012) & 0.054 (0.010) \\ 
  $f_{3}(x_3)$ & 1.00 & 0.016 (0.009) & 0.017 (0.009) & 0.021 (0.010) & 0.017 (0.012) \\ 
  \hline
    $f_{0}(x_0)$ & 1.60 & 0.054 (0.019) & 0.053 (0.019) & 0.055 (0.019) & 0.052 (0.022) \\ 
  $f_{1}(x_1)$ & 1.60 & 0.057 (0.020) & 0.056 (0.020) & 0.055 (0.018) & 0.046 (0.020) \\ 
  $f_{2}(x_2)$ & 1.60 & 0.089 (0.018) & 0.089 (0.019) & 0.089 (0.019) & 0.079 (0.018) \\ 
  $f_{3}(x_3)$ & 1.60 & 0.027 (0.015) & 0.027 (0.015) & 0.030 (0.016) & 0.029 (0.020) \\ 
\bottomrule
\end{tabular}
\cprotect\caption{Mean root-mean-square error of fitted terms in the case of equal sample sizes, unequal sample sizes, and mega-analysis, with residual standard deviation $\sigma=1.0$ or $\sigma=1.6$. Standard deviations across simulations are shown in parentheses.}
\label{tab:univariate_rmse}
\end{table}

\begin{table}
\small
\centering
\begin{tabular}{ll|cccc}
\toprule
Term & $\sigma$ & \thead{Equal sample \\ size} & \thead{Unequal sample \\ size} & \thead{Unequal range \\ and sample size} & \thead{Mega-analysis} \\
\hline
$f_{0}(x_0)$ & 1.00 & 0.95 (0.21) & 0.95 (0.21) & 0.95 (0.23) & 0.97 (0.16) \\ 
  $f_{1}(x_1)$ & 1.00 & 0.89 (0.31) & 0.90 (0.30) & 0.89 (0.31) & 0.97 (0.17) \\ 
  $f_{2}(x_2)$ & 1.00 & 0.88 (0.32) & 0.89 (0.31) & 0.87 (0.33) & 0.96 (0.20) \\ 
  $f_{3}(x_3)$ & 1.00 & 0.99 (0.10) & 0.99 (0.10) & 0.96 (0.19) & 0.99 (0.11) \\ 
  \hline
 $f_{0}(x_0)$ & 1.60 & 0.96 (0.20) & 0.96 (0.19) & 0.96 (0.20) & 0.98 (0.15) \\ 
  $f_{1}(x_1)$ & 1.60 & 0.86 (0.34) & 0.88 (0.33) & 0.91 (0.28) & 0.97 (0.17) \\ 
  $f_{2}(x_2)$ & 1.60 & 0.87 (0.33) & 0.87 (0.34) & 0.87 (0.34) & 0.96 (0.20) \\ 
  $f_{3}(x_3)$ & 1.60 & 0.99 (0.11) & 0.99 (0.11) & 0.98 (0.15) & 0.98 (0.13) \\ 
\bottomrule
\end{tabular}
\cprotect\caption{Mean coverage of 95 \% confidence intervals for fitted terms in the case of equal sample sizes, unequal sample sizes, and mega-analysis, with residual standard deviation $\sigma=1.0$ or $\sigma=1.6$. Standard deviations across simulations are shown in parentheses.}
\label{tab:univariate_coverage}
\end{table}

\subsection{Hypothesis Testing and Power}
\label{sec:power}

\begin{figure}
\center
\includegraphics[width=.8\linewidth]{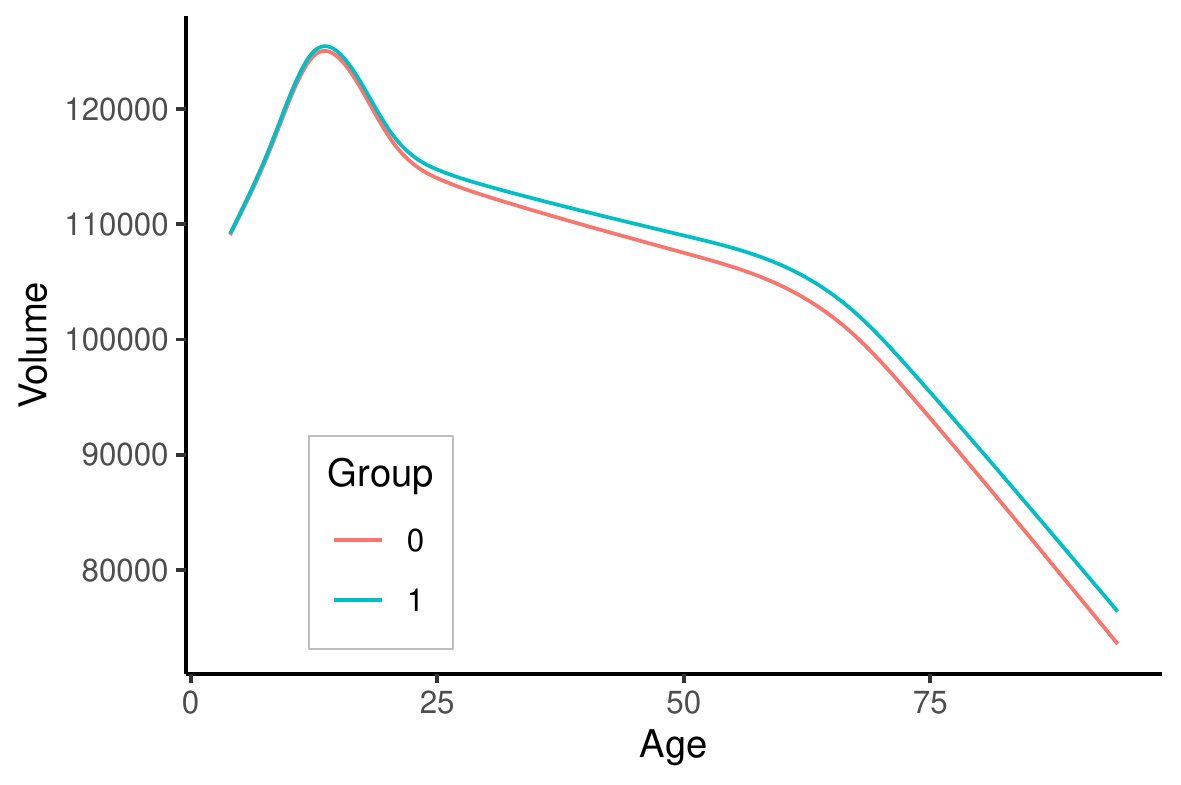}
\caption{\textbf{Lifespan trajectories with group interaction.} Functional forms assumed for lifespan trajectories in Section \ref{sec:power}. Subjects are assumed to belong to either group 0 or 1, whose mean lifespan trajectories differ as shown by the two curves.}
\label{fig:power_forms}
\end{figure}

A second set of simulation experiments was conducted with the goal of comparing the statistical inference performance of meta-analysis to mega-analysis. Two issues are of particular interest in this regard; first, whether the distribution of $p$-values is close to uniform when the null hypothesis is true (e.g., \citet{Murdoch2008}), and second, the power to reject a false null hypothesis. A nonlinear functional form approximating the lifespan trajectory of cerebellum cortex volume was estimated with the LCBC data \citep{Fjell2017,Walhovd2016}, as shown in Figure \ref{fig:power_forms}. For the power analysis, it was assumed that a dichotomous group variable interacted with the lifespan trajectory, leading to slightly higher atrophy for members of the baseline group, especially in advanced ages. For analysis of the null distribution of $p$-values, the two groups had identical lifespan trajectories. Analyzing this type of smooth interactions is relevant, e.g., when investigating the impact of a given genetic variation on lifespan trajectories of brain measures \citep{Walhovd2019}.

Cross-sectional measurements were simulated with age uniformly distributed between 4 and 94 years, and group memberships randomly allocated to 0 or 1 with equal probabilities. For the mega-analysis, all measurements were analyzed in a single GAM, while for the meta-analysis, the data were first split into 6 datasets and analyzed separately, before a meta-analytic $p$-value was computed. For reference, the power obtained when using a single dataset of size 1/6th of the total dataset was also computed. A total of 1,000 Monte Carlo samples were analyzed for each parameter setting. For the case of a nonzero group interaction, statistical power was computed as the fraction of the 1,000 random simulations in which the group interaction was significant at a 5 \% level. In the first set of simulations, the total sample size was fixed at 3,000 while the residual standard deviation varied between 1,000 and 15,000. In the second set of simulations, the residual standard deviation was fixed at 3,500, and the total sample size varied between 900 and 3,000. In all cases, "cohort fits" were computed by randomly splitting the dataset into 6 equally sized parts. The GAMs used to analyze the data in each sample were of the form
\begin{equation*}
y = \beta_{0,m} + f_{1,m}\left(x_{1}\right) + f_{2,m}\left(x_{1}\right) x_{2} + \beta_{2,m} x_{2} + \epsilon, ~ m = 1, \dots, M,
\end{equation*}
where $x_{1}$ is age, $x_{2}\in \{0, 1\}$ is an indicator for group membership, and $\epsilon$ is a normally distributed residual. The parameter $\beta_{0,m}$  represents the intercept, $\beta_{2,m}$ is the offset effect of membership in group 1, the smooth term $f_{1,m}(x_{1})$ represents the age trajectory of subjects in group 0, and $f_{2,m}(x_{1})$ represents the difference between the smooth term of subjects in group 1 and subjects in group 0. Hence, subjects in group 1 have age trajectory given by $f_{1,m}\left(x_{1}\right) + f_{2,m}\left(x_{1}\right)$. GAMs were fitted with the \verb!gam! function in \verb!mgcv! \citep{Wood2017}, using cubic regression splines to construct the smooth terms and generalized cross-validation for smoothing. Knot placement was determined independently for each study. The null hypothesis states that there is no difference between the lifespan trajectories across groups, and the $p$-values corresponding to this null hypothesis in each sample were directly obtained from the model fit, which uses the methods described in \citet{Wood2012}. For the meta-analysis, we compared several different methods for combining $p$-values: Wilkinson's maximum $p$ \citep{Wilkinson1951}, Tippet's minimum $p$ \citep{Tippet1931}, the logit-$p$ method \citep{Becker1994}, Fisher's sum of logs \citep{Fisher1925}, Edgington's sum of $p$ \citep{Edgington1972}, and Stouffer's sum of z \citep{Stouffer1949}, using the implementations in the R package \verb!metap! \citep{Dewey2019}. As all samples in the meta-analysis were of equal size, equal meta-analytic weights were used in Stouffer's sum of z \eqref{eq:Stouffer}. The other methods do not use weights. Tippet's minimum $p$ method gave $p$-values closest to uniform under the null hypothesis under most parameter settings, while Stouffer's sum of z method typically gave highest power. The $p$-values resulting from these two methods are hence shown in the results in this section, while complete results for all methods can be found in Supplementary Material III.

Figure \ref{fig:null_curves} shows quantile-quantile plots of the $p$-values obtained by meta-analysis, mega-analysis, and a fit of a single dataset in the case of no actual interaction between the group variable and the lifespan trajectories in the case with sample size 3,000 and residual standard deviation 3,500. Results for other values of these parameters were similar, and are shown in Supplementary Material III. The gray line shows the ideal reference line. All methods yielded $p$-values which deviated to some degree from the uniform distribution. Meta-analytic $p$-values computed using Tippet's minimum $p$ method were close to the $p$-values obtained either in the mega-analysis or in the single data fit. $p$-values computed using Stouffer's sum of z, on the other hand, were considerably further from being uniformly distributed. As Figure \ref{fig:null_curves} shows, the $p$-values of the mega-analysis were not perfectly uniformly distributed. This is due to the approximate nature of the algorithms used to compute $p$-values in GAMMs, which need to take into account the overall uncertainty in the smoothing parameter \citep[Sec. 6.12]{Wood2017}.

\begin{figure}
\includegraphics[width=.8\columnwidth]{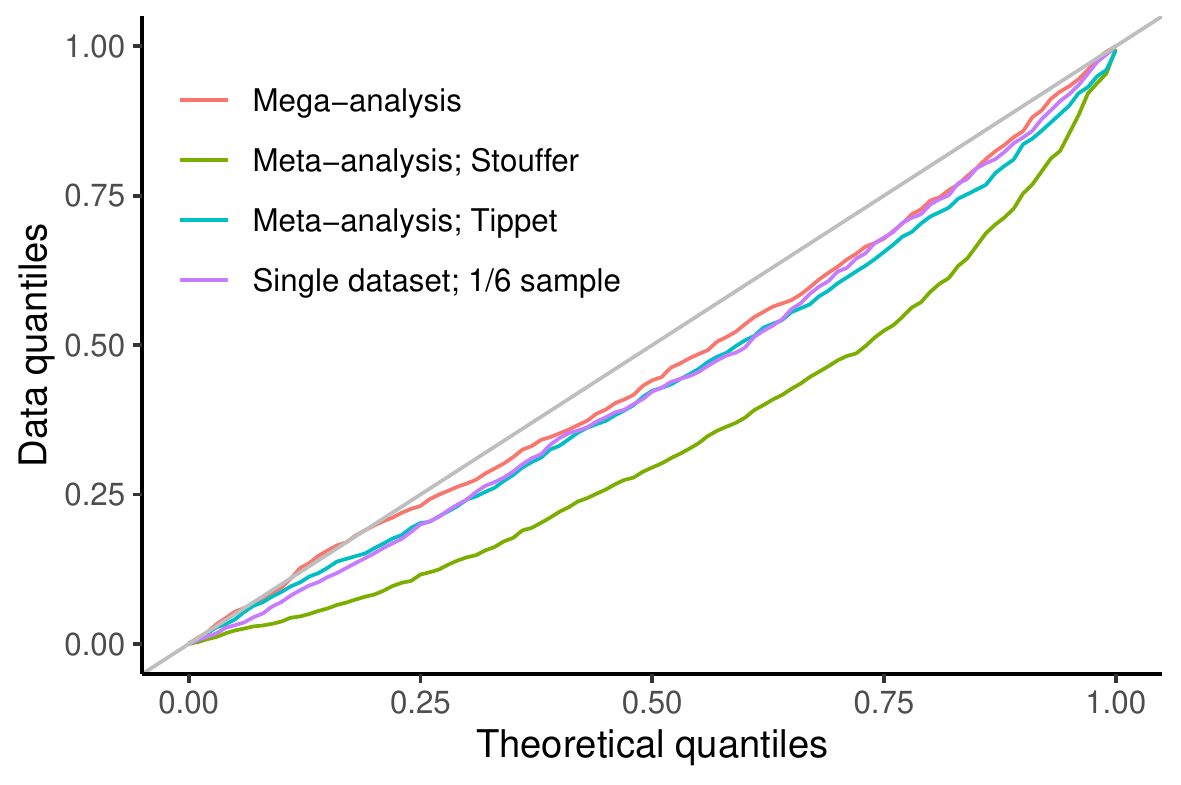}
\cprotect\caption{\textbf{P-value distribution under the null hypothesis.} Quantile-quantile plot of $p$-values under the null hypothesis as described in Section \ref{sec:power}, for the case of residual standard deviation equal to 3,500 and total sample size 3,000. Meta-analytic $p$-values were computed using both Stouffer's and Tippet's method, as shown by the legend.}
\label{fig:null_curves}
\end{figure}

\begin{figure}
\includegraphics[width=\columnwidth]{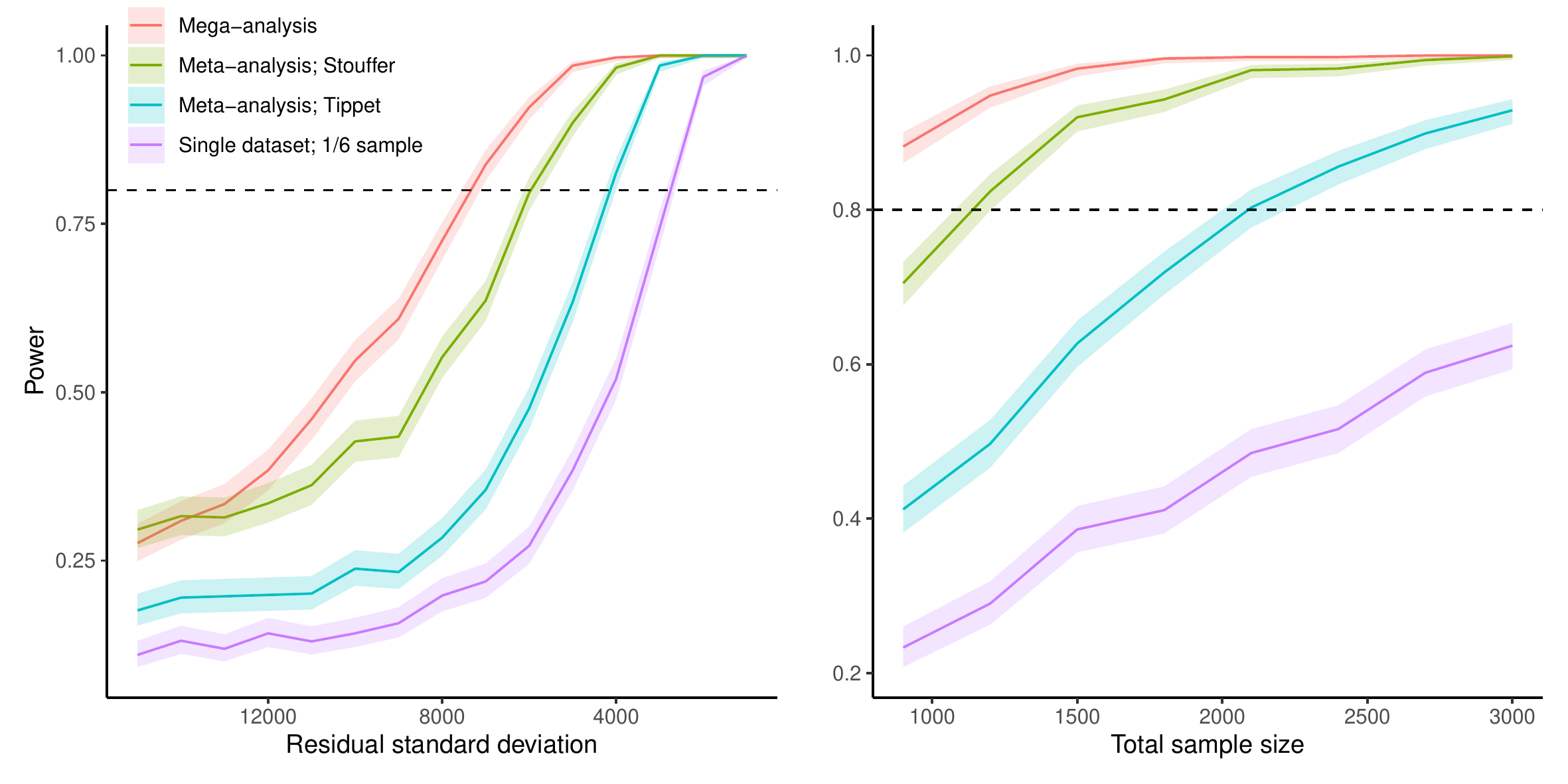}
\cprotect\caption{\textbf{Statistical power to detect interaction.} Results of statistical power simulations desribed in Section \ref{sec:power}. Left: fixed total sample size 3,000 and varying noise level. Right: fixed noise level $\sigma$=3,500 and varying total sample size. Shaded areas around curves show 95 \% confidence intervals computed using the R package \verb!Hmisc! \citep{Harrel2019}. Meta-analytic $p$-values were computed using both Stouffer's and Tippet's method, as shown by the legend.}
\label{fig:power_curves}
\end{figure}

Figure \ref{fig:power_curves} (left) shows power curves for varying residual standard errors, and Figure \ref{fig:power_curves} (right) shows power curves over a range of sample sizes. In both cases, the mega-analytic approach outperforms the meta-analytic approaches. Stouffer's sum of z method obtained power closest to the mega-analysis, while Tippet's minimum p method had lower power. Analyzing a single dataset, representing 1/6th of the total data, gave much lower power than either of the other two approaches. This highlights the benefit of pointwise meta-analysis compared to separate analyses by each center, when data cannot be shared.

To summarize, meta-analysis using Stouffer's sum of z method had power fairly close to that of a mega-analysis, at an increased risk of falsely rejecting true null hypotheses. On the other hand, meta-analysis using Tippet's minimum $p$ method had risk of falsely rejecting a true null hypothesis close to that of a mega-analysis, at the cost of lower power. The other methods for combining $p$-values were somewhere inbetween these extremes, as shown in Supplementary Material III.

\section{Case Study}
\label{sec:case_study}

We will now illustrate the proposed methods on brain imaging data from six European cohorts analyzed by \citet{Fjell2019}. The datasets contained measurements of sleep quality and hippocampal volume from the Berlin Study of Aging-II (BASE-II) \citep{Bertram2013,Gerstorf2016}, the Betula project \citep{Nilsson1997}, the Cambridge Centre for Ageing and Neuroscience study (Cam-CAN) \citep{Taylor2017}, Center for Lifespan Changes in Brain and Cognition longitudinal (LCBC) studies \citep{Walhovd2016,Fjell2017}, Whitehall-II \citep{Filippini2014}, and University of Barcelona brain studies \citep{AbellanedaPrez2019,Rajaram2017,VidalPineiro2014}. Self-reported sleep and hippocampal volume data from 2,843 participants (18-90 years) were included. Longitudinal information on hippocampal volume was available for 1,065 participants, yielding a total of 4,621 observations. Mean interval from first to last examination was 3.8 years (range 0.2-11.0 years). Participants were screened to be cognitively healthy and in general not suffer from conditions known to affect brain function, such as dementia, major stroke, multiple sclerosis, etc. Exact screening criteria were not identical across subsamples. Detailed sample characteristics are presented in the Supplementary Material I.

In \citet{Fjell2019}, the data were analyzed jointly using GAMMs in a mega-analysis, taking into account both the clustering of repeated measurements within the same subject, and of subjects within a given cohort. However, the methods proposed in this paper enable this type of multi-cohort analysis also when the data cannot be shared. In this particular example we examine how hippocampal volume is related to age and to sleep quality as measured by the global score on the Pittsburgh Sleep Quality Index (PSQI) \citep{Buysse1989}. A low value of the PSQI variable indicates good sleep.

\begin{figure}
\centering
\includegraphics[width=.8\linewidth]{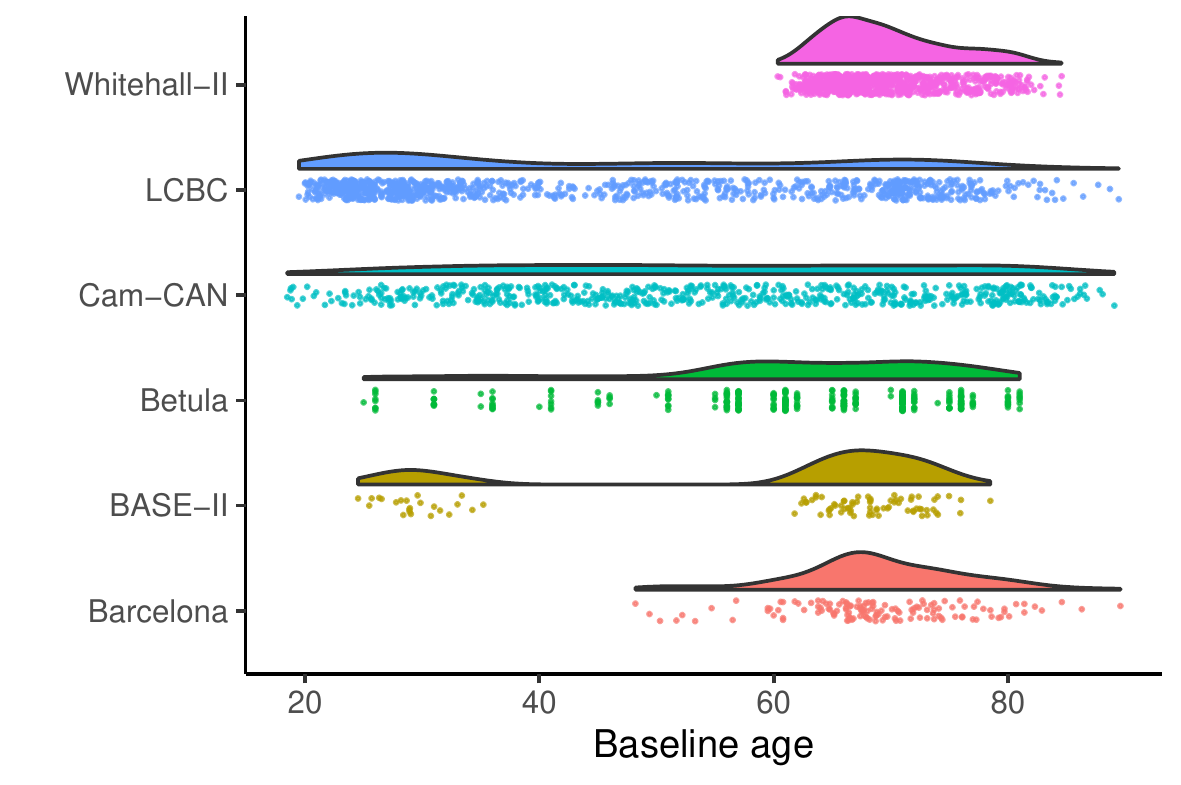}
\includegraphics[width=.8\linewidth]{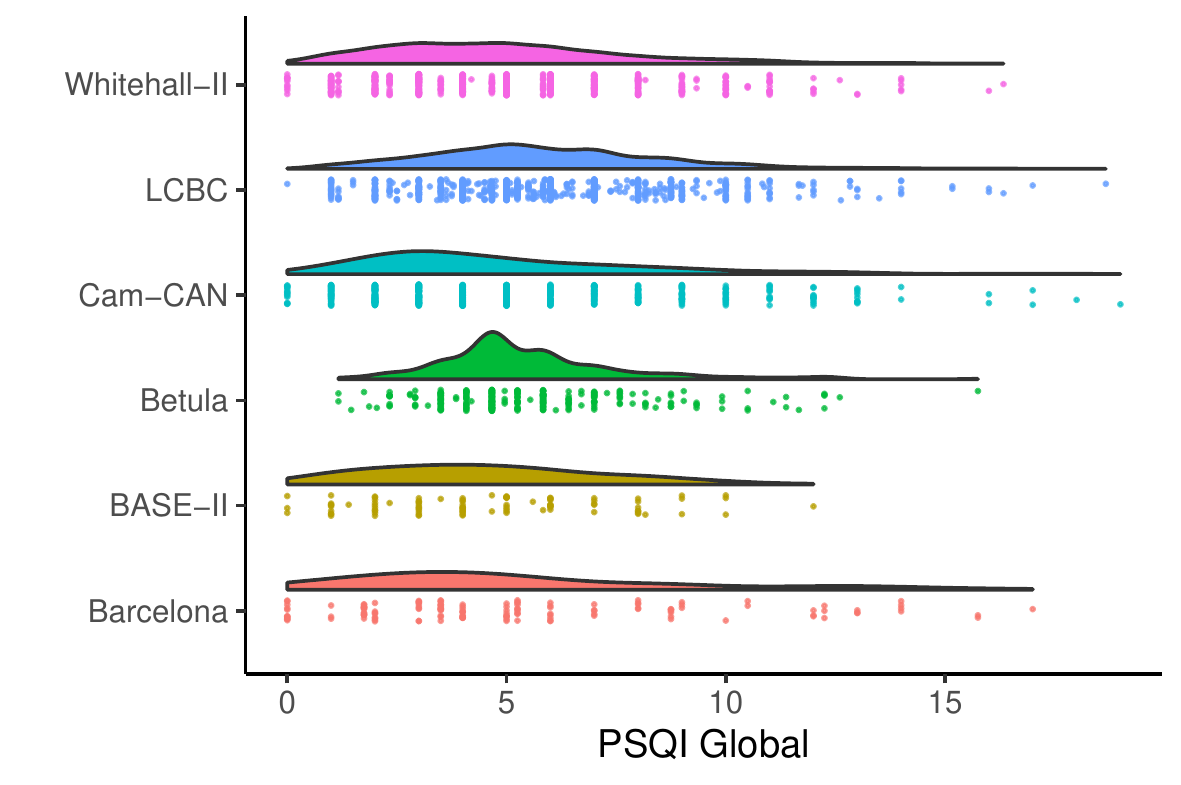}
\caption{\textbf{Empirical distribution of explanatory variables.} Raincloud plots \citep{Allen2019} showing the distribution of baseline age (top) and global PSQI score (bottom) in the data from each study in Section \ref{sec:case_study}.}
\label{fig:cohort_dist}
\end{figure}

\begin{figure}
\centering
\includegraphics[width=.8\linewidth]{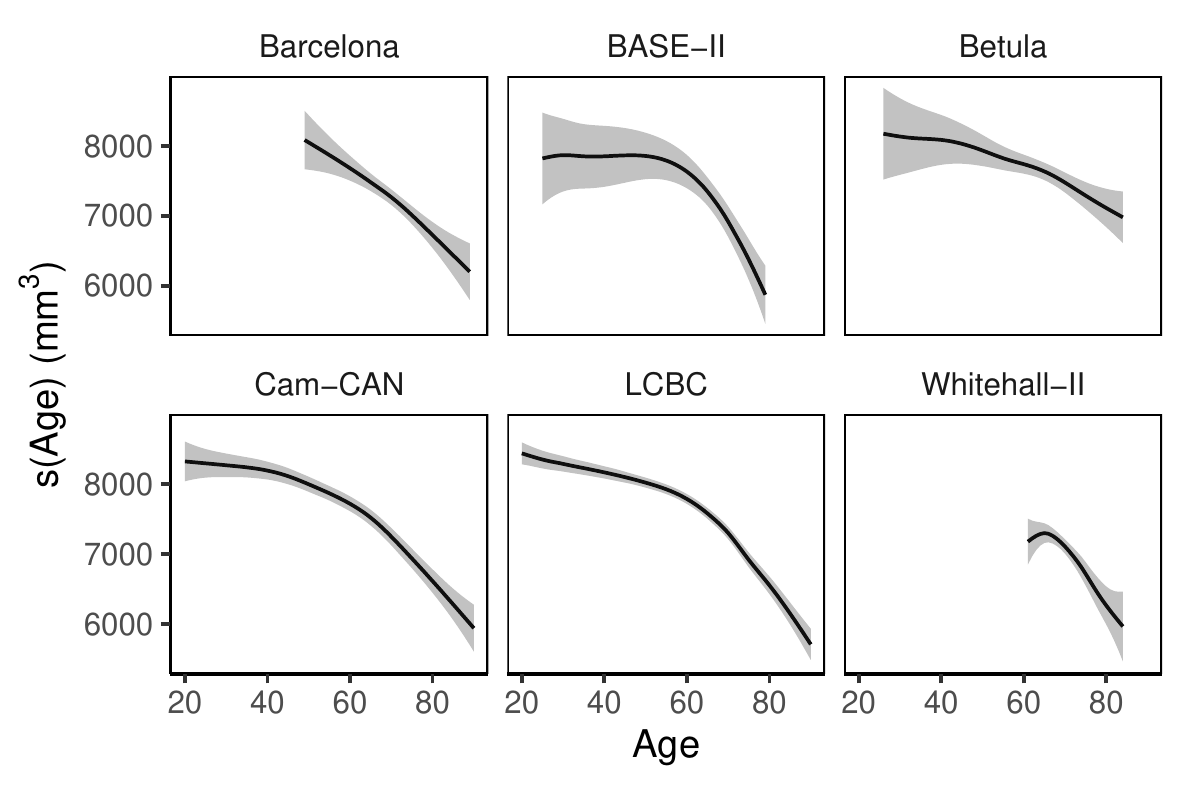}
\caption{\textbf{Age trajectories for each cohort.} Estimates of $\beta_{0} + f_{1}(x_{1})$ in \eqref{eq:cohort_model}, showing how age predicts hippocampal volume in each cohort. Gray shaded areas are 95 \% confidence intervals.}
\label{fig:cohort_age_fits}
\end{figure}

The following model was first fit to data from each study separately:
\begin{equation}
\label{eq:cohort_model}
y_{ij} = \beta_{0} + f_{1}(x_{ij,1}) + f_{2}(x_{ij,1})x_{i,2} + \beta_{3} x_{i,3} + b_{i} + \epsilon_{ij}.
\end{equation}
$y_{ij}$ denotes hippocampal volume of subject $i$ at timepoint $j$, $x_{ij,1}$ is the age of subject $i$ at timepoint $j$, $x_{i,2}$ is the global PSQI score, and $x_{i,3}$ is the sex of subject $i$. $b_{i} \sim N(0, \sigma_{b}^{2})$ is the random intercept of subject $i$ and $\epsilon_{ij} \sim N(0, \sigma^{2})$ is the residual. The main effect of age is represented by $f_{1}(x_{1})$. $f_{2}(x_{1})x_{2}$ is a varying-coefficient term \citep{Hastie1993}, in which $f_{2}(x_{1})$ is a regression coefficient for sleep quality which varies smoothly with age. Restricted maximum likelihood was used both for smoothing and estimation of random effect terms, and cubic regression splines were used as basis functions. The range of the age variable differed considerably between studies, as shown in the top part of Figure \ref{fig:cohort_dist}. Hence, both the knot placement and the number of knots used to fit $f_{1}(x_{1})$ and $f_{2}(x_{1})$ was determined for each cohort separately. The simulation procedure described in \citet[Ch. 5.9]{Wood2017} was used to ensure that the number of knots was large enough to allow sufficient flexibility for the shapes of the smooth terms. The sleep quality scores were similarly distributed across cohorts, as shown in the bottom part of Figure \ref{fig:cohort_dist}. Betula differs somewhat in shape from the others, due to a transformation that had to be applied to these data \citep{Fjell2019}. Figure \ref{fig:cohort_age_fits} shows the fits of the term $\beta_{0} + f_{1}(x_{1})$ in \eqref{eq:cohort_model} relating age to hippocampal volume, over the range of ages in each cohort. 

For the meta-analysis, we will focus on the effect of age on hippocampal volume including the intercept term, $\beta_{0} + f_{1}(x_{1})$, and the age-dependent effect of sleep quality on hippocampal volume, $f_{2}(x_{1})$. To this end, we set up a grid over which to compute the estimates, containing the range of ages from 20 to 90 equally spaced by 0.1 year, and the value of the sleep quality score set to $x_{2}=1$, such that $f_{2}(x_{1}) x_{2} = f_{2}(x_{1})$, representing the main effect of sleep as a function of age. Random effects meta-analysis was used, with between-study variance estimated with the DerSimonian-Laird estimator shown in equation \eqref{eq:between_var_estimator}.

\begin{figure}
\centering
\includegraphics[width=\linewidth]{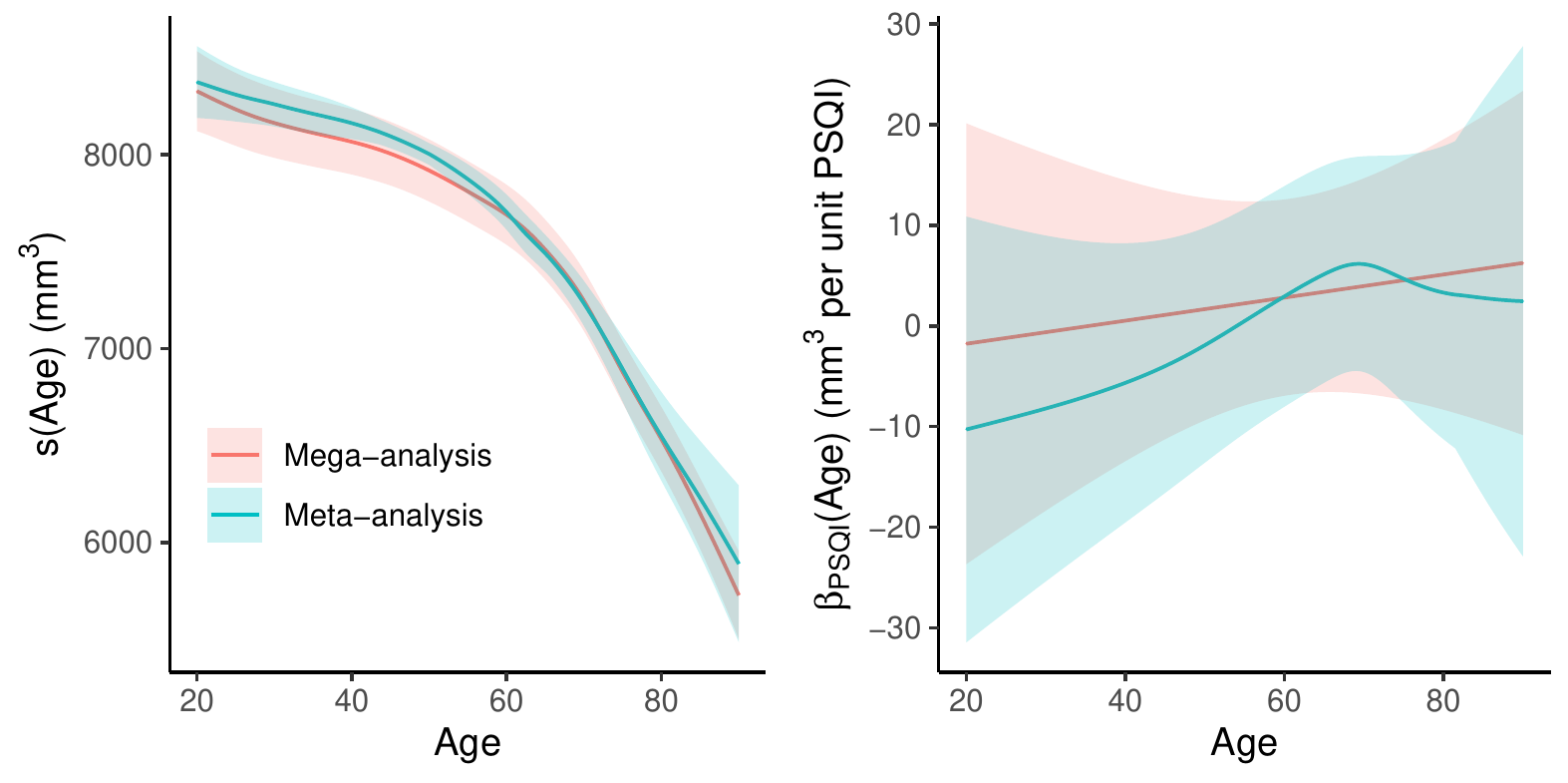}
\caption{\textbf{Comparison of meta-analytic and mega-analytic estimates.} Meta-analytic fits obtained as described in Section \ref{sec:case_study}, compared to the corresponding fit obtained with full data. Left: effect of age on hippocampal volume, including the overall intercept. Right: effect of PSQI global score on hippocampal volume as a function of age.}
\label{fig:meta_fits}
\end{figure}

\begin{figure}
\centering
\includegraphics[width=\linewidth]{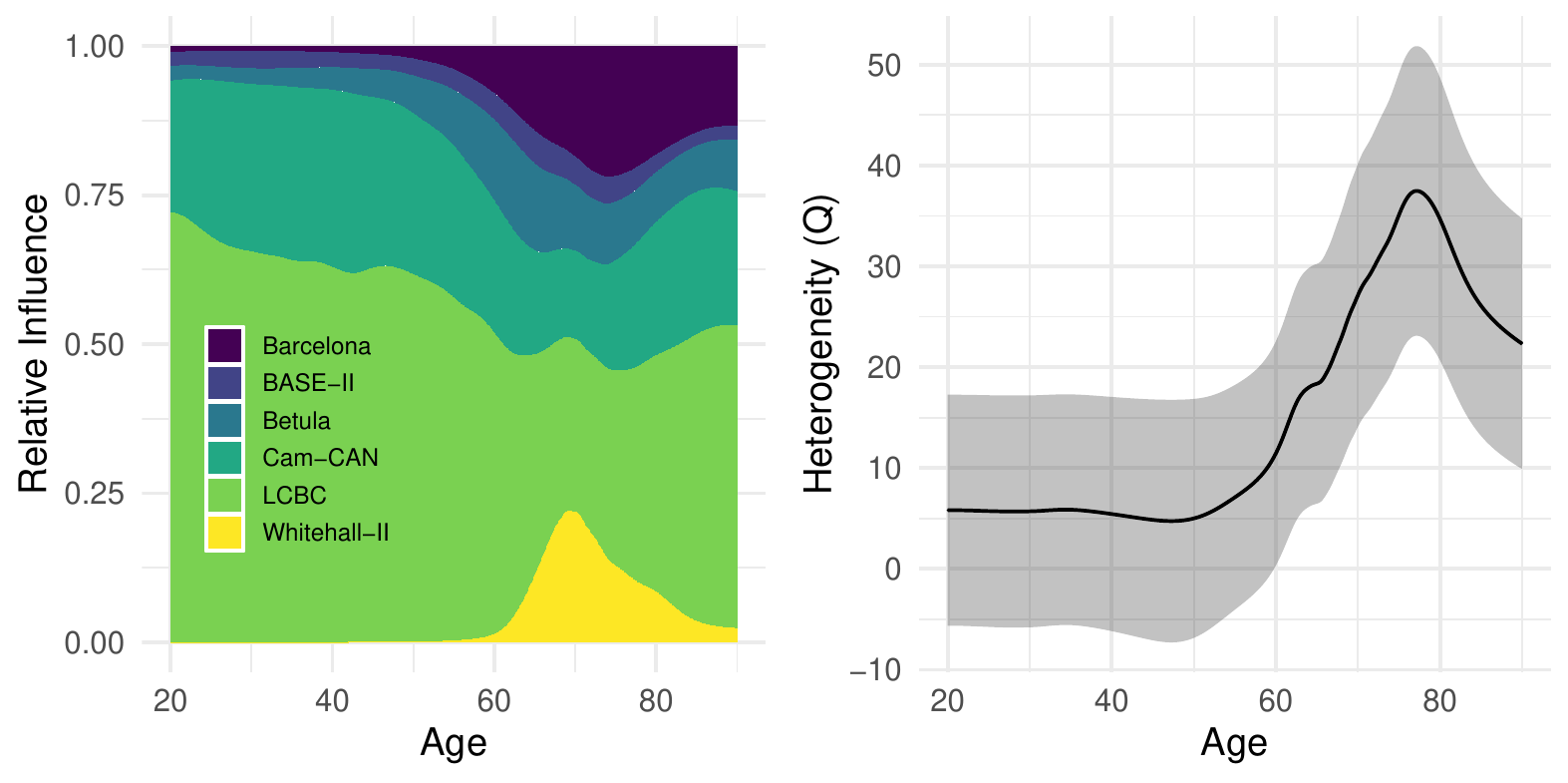}
\caption{\textbf{Dominance and heterogeneity plots.} Dominance and heterogeneity plots for $\beta_{0} + f_{1}(x_{1})$ in equation \eqref{eq:cohort_model}. Left: the relative contribution from each study to the meta-analytic fit over age. Right: Cochran's Q statistic for heterogeneity over age. Shaded areas represent 95 \% confidence intervals.}
\label{fig:dominance_plot}
\end{figure}

Figure \ref{fig:meta_fits} shows the meta-analytic fits compared to the full data case. The estimated effects of age on hippocampal volume are very similar between the two approaches, although the meta-analytic fit lies somewhat above the mega-analytic fit for ages below 60 and has somewhat narrower confidence bands at low ages and wider confidence bands at high ages. A possible reason for the narrow confidence bands of the meta-analytic estimate of $f_{1}(x_{1})$ for ages in the range from 30 to 55 years is that this age range is dominated by LCBC and Cam-CAN (Figure \ref{fig:dominance_plot}), which have very similar estimated functional forms (Figure \ref{fig:cohort_age_fits}). As shown in Supplementary Material IV (p. 16), the estimated between-sample variance is even identically zero over part of this range. Since the standard error of the meta-analytic fit is estimated independently at each age (cf. equation \eqref{eq:meta_se}), the confidence bands hence become narrow, in contrast to the mega-analytic fit, for which the global smoothness assumption and the utilization of repeated measurements contribute to confidence bands whose width has little variation in the interior of the age range.

As in \citet{Fjell2019}, there seems to be no effect of global PSQI score on hippocampal volume at any age, as can be seen by the confidence intervals covering zero in both cases (Figure \ref{fig:meta_fits}, right). In the meta-analytic case, the estimated curve has a peak at around 70 years, as opposed to the straight line estimated by the full data analysis. However, the confidence bands obtained with the two methods are highly overlapping. We note that while the mega-analysis estimates a linear varying-coefficient term $f_{2}(x_{1})$, the meta-analytic estimate is nonlinear. As shown in Supplementary Material IV, all the individual cohort fits except Betula were very close to linearity. However, pointwise meta-analytic fits are nonlinear by construction, so even if all individual cohort fits estimated a linear effect, the meta-analytic estimate would in general be nonlinear. This can be seen by the fact that $\hat{f}_{s}(\mathbf{x})$ depends nonlinearly on the covariates $\mathbf{x}$ in equation \eqref{eq:meta_est}, through the products of the estimated smooth terms with the meta-analytic weights. In contrast, the mega-analysis shrinks the total estimate towards a linear function through the second-derivative penalty. As a result, the mega-analytic estimate will be linear when the data do provide sufficient evidence of a nonlinear effect.

In order to quantify how much each study contributes to the meta-analytic fit at each value of an explanatory variable, we propose using dominance plots, visualizing $\hat{\sigma}_{s,m}^{2}/\text{se}_{\hat{f}_{s}}^{2}$ for $m=1,\dots,M$. Figure \ref{fig:dominance_plot} (left) shows that LCBC and Cam-CAN are the main contributors to the meta-analytic fit for the main effect of age on hippocampal volume for ages up to around 50 years, after which the relative influence of the other studies starts increasing. Furthermore, the heterogeneity of the models fit in each study can be analyzed by computing Cochran's Q statistic \citep{Cochran1954} over an explanatory variable, thus comparing $\hat{f}_{s,m}$ for $m=1,\dots,M$ independently at each value of the explanatory variable. Figure \ref{fig:dominance_plot} (right) shows a heterogeneity plot comparing the main effects of age in each study, with 95 \% confidence intervals represented by the shaded gray areas. The confidence interval in the heterogeneity plot does not contain zero for ages above 60, indicating that there is evidence of systematic differences across cohorts in the effect of age on hippocampal volume after the age of 60.

\subsection{Pointwise Meta-Analysis in R with the 'metagam' Package}
\label{sec:metagam}

This section shows how the meta-analysis described above can be conducted in R using the \verb!metagam! package, which implements the methods presented in this paper. Some details are omitted for clarity, and are shown in Supplementary Material IV.

First, the following code fits a GAMM to the data for each study using the \verb!mgcv! package \citep{Wood2017}.

\begin{knitrout}
\definecolor{shadecolor}{rgb}{0.969, 0.969, 0.969}\color{fgcolor}\begin{kframe}
\begin{alltt}
\hlkwd{library}\hlstd{(mgcv)}
\hlcom{# Fit GAMM in cohort 1}
\hlstd{cohort_gam1} \hlkwb{<-} \hlkwd{gamm}\hlstd{(}
  \hlstd{Hippocampus} \hlopt{~} \hlkwd{s}\hlstd{(Age)} \hlopt{+} \hlkwd{s}\hlstd{(Age,} \hlkwc{by} \hlstd{= PSQI_Global)} \hlopt{+} \hlstd{Sex,}
  \hlkwc{data} \hlstd{= cohort_data1,} \hlkwc{random} \hlstd{=} \hlkwd{list}\hlstd{(}\hlkwc{ID} \hlstd{=}\hlopt{~} \hlnum{1}\hlstd{),} \hlkwc{method} \hlstd{=} \hlstr{"REML"}\hlstd{)}
\end{alltt}
\end{kframe}
\end{knitrout}

The fitted model objects returned by \verb!gamm()! contain the original data used to fit the model, as well as the responses. The \verb!strip_rawdata()! function from \verb!metagam! removes all individual participant data from each model fit, returning an object containing only aggregate quantities that can be shared without any individual data. The following lines attach the \verb!metagam! package and then create an object \verb!cohort_fit1!, which does not contain any individual-specific data.

\begin{knitrout}
\definecolor{shadecolor}{rgb}{0.969, 0.969, 0.969}\color{fgcolor}\begin{kframe}
\begin{alltt}
\hlkwd{library}\hlstd{(metagam)}
\hlstd{cohort_fit1} \hlkwb{<-} \hlkwd{strip_rawdata}\hlstd{(cohort_gam1)}
\end{alltt}
\end{kframe}
\end{knitrout}

Assuming each cohort has followed the two steps above, the following code gathers the model fits from each of the six cohorts in a list, creates a grid over which to predict, and finally uses the \verb!metagam()! function to compute the meta-analytic fits. 

\begin{knitrout}
\definecolor{shadecolor}{rgb}{0.969, 0.969, 0.969}\color{fgcolor}\begin{kframe}
\begin{alltt}
\hlcom{# Combine fits from each cohort in a list}
\hlstd{cohort_fits} \hlkwb{<-} \hlkwd{list}\hlstd{(cohort_fit1, cohort_fit2, cohort_fit3,}
                    \hlstd{cohort_fit4, cohort_fit5, cohort_fit6)}

\hlcom{# Create a grid over which to compute meta-analytic fits}
\hlstd{grid} \hlkwb{<-} \hlkwd{data.frame}\hlstd{(}
  \hlkwc{Age} \hlstd{=} \hlkwd{seq}\hlstd{(}\hlkwc{from} \hlstd{=} \hlnum{20}\hlstd{,} \hlkwc{to} \hlstd{=} \hlnum{90}\hlstd{,} \hlkwc{by} \hlstd{=} \hlnum{0.1}\hlstd{),}
  \hlkwc{Sex} \hlstd{=} \hlkwd{factor}\hlstd{(}\hlstr{"Female"}\hlstd{,} \hlkwc{levels} \hlstd{=} \hlkwd{c}\hlstd{(}\hlstr{"Female"}\hlstd{,} \hlstr{"Male"}\hlstd{)),}
  \hlkwc{PSQI_Global} \hlstd{=} \hlnum{1}\hlstd{)}

\hlcom{# Smooth function of x_1, including overall intercept}
\hlstd{metafit_age} \hlkwb{<-} \hlkwd{metagam}\hlstd{(cohort_fits, grid,} \hlkwc{terms} \hlstd{=} \hlstr{"s(Age)"}\hlstd{,}
                       \hlkwc{method} \hlstd{=} \hlstr{"DL"}\hlstd{,} \hlkwc{intercept} \hlstd{=} \hlnum{TRUE}\hlstd{)}

\hlcom{# Age-varying slope of x_2, not including overall intercept}
\hlstd{metafit_psqi} \hlkwb{<-} \hlkwd{metagam}\hlstd{(cohort_fits, grid,}
                        \hlkwc{terms} \hlstd{=} \hlstr{"s(Age):PSQI_Global"}\hlstd{,}
                        \hlkwc{method} \hlstd{=} \hlstr{"DL"}\hlstd{,} \hlkwc{intercept} \hlstd{=} \hlnum{FALSE}\hlstd{)}
\end{alltt}
\end{kframe}
\end{knitrout}

The argument \verb!method = "DL"! specifies that random effects meta-analysis should be used, with the DerSimonian-Laird estimator \citep{derSimonian1986}. The \verb!metafor! package \citep{Viechtbauer2010} performs the actual estimation, and all estimators available in \verb!metafor! may be used. By default, predictions from each model are computed over the whole supplied grid, thus extrapolating the estimates from cohorts whose data cover only a subset of the grid. Arguments can be specified in order to compute the predictions from each model only within the range of variables used to fit it. In practice, this latter option does not have much impact, since the standard errors are large outside of the range of the variables used in the fit, and hence the corresponding predictions get a very low weight at these points.

Finally, the dominance and heterogeneity plots shown in Figure \ref{fig:dominance_plot} are obtained with the commands:

\begin{knitrout}
\definecolor{shadecolor}{rgb}{0.969, 0.969, 0.969}\color{fgcolor}\begin{kframe}
\begin{alltt}
\hlkwd{plot_dominance}\hlstd{(metafit_age)}
\hlkwd{plot_heterogeneity}\hlstd{(metafit_age)}
\end{alltt}
\end{kframe}
\end{knitrout}

\section{Discussion}

We have proposed and illustrated a flexible way to obtain meta-analytic fits of GAMs in neuroimaging studies where individual participant data cannot be shared across cohorts. In the simulation studies, the meta-analytic procedure showed estimation performance close to that obtained in the ideal case, in which all data were analyzed in a single model, except that the meta-analytic estimates tended to have somewhat too narrow confidence intervals. Furthermore, the simulations showed that when testing for an interaction between a smooth function and a categorical variable, the distribution of $p$-values under the null hypothesis of no interaction, and the power to detect an actual interaction, were highly dependent on the chosen method for combining $p$-values, offering a trade-off between power and the probability of making false rejections. The proposed method is particularly useful when the variables under study have different ranges across cohorts, such that enforcing the same knot placement is suboptimal and might lead to nonidentified models. This is the case in many multi-cohort and consortium studies using neuroimaging data, where for instance age-range or patient distribution across a clinical indicator may vary considerably across samples. Differing variable ranges and knot placement are also inevitable across independent studies using GAMs to estimate some effect of interest in different study populations.

A case study illustrating the use of pointwise meta-analysis was considered in Section \ref{sec:case_study}, in which the effect of sleep quality and age on hippocampal volume was estimated for six European cohorts. Due to the nonlinear lifespan relationship between age and hippocampal volume, GAMMs were preferable to LMMs when analyzing these data. However, the highly varying age distributions (Figure \ref{fig:cohort_dist}) lead to nonidentified models when the same knot location was enforced across cohorts (cf. Table \ref{tab:equal_knots_meta}). Meta-analysis of GAMMs by combining spline weights at each knot \citep{Gasparrini2012} could hence not be used. The pointwise meta-analysis developed in this paper alleviated these issues, and allowed computing meta-analytic estimates of both the effect of age on hippocampal volume and the age-varying effect of sleep quality on hippocampal volume. Since the full data were available in a single location in this case, the meta-analytic estimates could be directly compared to a mega-analysis in which all data were analyzed jointly. The meta-analytic estimate of the effect of age on hippocampal volume was very close to the mega-analytic estimate (Figure \ref{fig:meta_fits}, left), although it had slightly narrower confidence bands for the middle age ranges. The meta-analytic estimate of the effect of sleep was also close to the mega-analytic estimate, both being almost zero over the full age range. A notable difference in the latter case was that while the mega-analysis estimated the effect of sleep to vary linearly with age, the meta-analytic estimate was nonlinear, as it will be by construction. An interesting topic for further study, which would enable a meta-analytic estimate to be linear when the smooth terms from each cohort are close to linear, involves imposing additional constraints on the meta-analytic fit, by using the degrees of freedom of the estimate from each cohort to inform the shape of the overall meta-analytic estimate. Dominance and heterogeneity plots (Figure \ref{fig:dominance_plot}) were also introduced as additional tools for analyzing the relative impact of each dataset on the meta-analytic fit, and the heterogeneity of the estimated effects, respectively, both as functions of age. 

One particular area of application for meta-GAM is imaging genetics. The need for very large sample sizes has long been recognized \citep{Thompson2014}, which imposes challenges due to privacy and data protection as well as practical issues regarding transfer, storage and processing of large amounts of neuroimaging data. These challenges have successfully been overcome in initiatives such as ENIGMA \citep{Bearden2017,Thompson2017} using a meta-analytic approach to gene discovery. Classic meta-analytic techniques are often inappropriate in situations where genetic effects are studied in interaction with other variables, such as age in a lifespan study. To test whether effects of genetic variants on a neuroimaging outcome measure vary as a function of age, or whether the lifespan trajectories of a neuroimaging outcome variable differ as a function of genetic variation \citep{Piers2018,Walhovd2019}, more complex modeling is needed. This functionality is provided by meta-GAM. As shown in Figure 8, this meta-analytic approach yielded superior power to detect effects in such situations compared to single studies, although not completely reaching the same statistical power as mega-analyses in cases of total sample size less than 2,000. Other examples of situations where meta-GAM would be applicable are when testing whether an effect varies as a function of another continuous variable, such as blood pressure, BMI or sleep duration. In all of these cases, the neuroanatomical outcome variable is expected to show a more complex relationship to the predictor variable than what can be captured by a parametric model. In these cases, meta-analytic GAM will be a powerful strategy to test genetic effects. Thus, we believe the present strategy may be a useful tool in neuroimaging genetics.

An alternative to the pointwise meta-analysis approach presented in this paper is to treat the fitted smooth functions from each cohort as samples from a Gaussian process \citep[Ch. 15]{Murphy2012}. A meta-analytic fit could then be obtained by using these samples to estimate the parameters of a common smoothing kernel. This approach has been taken by \citet{SalimiKhorshidi2011} for meta-analysis of neuroimaging data. Another alternative is using multiple imputation methods to generate synthetic data in each cohort with the same distributional properties as the original data, which can then be shared and analyzed in a mega-analysis \citep{Little1993,Rubin1993,Nowok2016}. Other possible extensions include accommodating potential correlation between the pointwise estimates in a given cohort using the robust variance estimation methods developed by \citet{Hedges2010}, and to model the effect of cohort-specific covariates using multivariate meta-regression \citep{Berkey1998}. The latter may be used to account for systematic differences between trajectories across cohorts (cf. Figure \ref{fig:dominance_plot}, right), and hence reduce potential bias in the meta-analytic estimates \citep{Hofer2009}. Also, deriving meta-analytic weights to use when combining $p$-values \citep{Rosenthal1978} as in Section \ref{sec:power} could potentially yield $p$-values closer to those of the mega-analysis.

Although we have focused on the case in which data are not available in a single location, the proposed methods can also be useful in two-stage analysis with GAMs. In two-stage analysis, models are fitted separately for each cohort as described here, and then fit using meta-analytic techniques \citep{Burke2016}. This approach seems to be somewhat less efficient than analyzing the data jointly in a one-stage model \citep{Boedhoe2019,Kontopantelis2018}, but is useful when combining the data is impractical due to storage requirements or harmonization challenges \citep{Sung2014}. Finally, use of meta-GAM as a research synthesis method requires estimates and covariance matrices of spline weights as well as knot placement and basis functions to be properly reported by the studies to be combined in the meta-analysis. The \verb!metagam! package easily allows extraction of such parameters from GAMs, creating model objects which can be made publicly available in repositories like the Open Science Framework \citep[https://osf.io/]{FosterMSLS2017}.

\section{Conclusion}

Here we propose and demonstrate an approach to meta-analysis of neuroimaging results in situations where parametric models might not be appropriate, such as is often the case, e.g., in lifespan research. Parametric models might not be able accurately to capture lifespan trajectories of most neuroanatomical volumes, here as demonstrated for hippocampus. We show how such data can be analyzed using meta-analysis of generalized additive (mixed) models, and demonstrate that this is a powerful approach using simulated as well as real multi-cohort longitudinal data from the Lifebrain consortium. We believe this approach can be successfully applied in a range of settings where neuroimaging variables are used as outcome, especially within lifespan and neuroimaging genetics research, and beyond.

\section*{Declaration of Competing Interest}

The authors declare that they have no competing interests.

\section*{Data and Code Availability Statement}

The R scripts used to conduct the simulation studies in Section \ref{sec:simsec} are available in Supplementary Material V, and the R script used in the case study in Section \ref{sec:case_study} are available in Supplementary Material IV. The R package \verb!metagam! implementing the methods developed in this paper is available from the Comprehensive R Archive Network (https://cran.r-project.org/package=metagam).

The data supporting the results of the current study are available from the corresponding author on reasonable request, given appropriate ethical and data protection approvals. Requests for data included in the Lifebrain meta-analysis can be submitted to the relevant principal investigators of each study. Contact information can be obtained from the corresponding author.

\section*{Ethics Statement}

The Lifebrain project is approved by the Regional Committee for Medical and Health Research Ethics of South Norway. Each sub-study was approved by the relevant ethical review board in the respective country (see Supplementary Material I).

\section*{Acknowledgement}

The Lifebrain project is funded by the EU Horizon 2020 Grant: ‘Healthy minds 0–100 years: Optimising the use of European brain imaging cohorts (“Lifebrain”)’. Grant agreement number: 732592. Call: Societal challenges: Health, demographic change and well-being. In addition, the different sub-studies are supported by different sources:
LCBC: The European Research Council under grant agreements 283634, 725025 (to A.M.F.) and 313440 (to K.B.W.), as well as the Norwegian Research Council (to A.M.F., K.B.W.). Betula: a scholar grant from the Knut and Alice Wallenberg (KAW) foundation to L.N. University of Barcelona: Partially supported by a Spanish Ministry of Economy and Competitiveness (MINECO) grant to D-BF $[$grant number PSI2015-64227-R (AEI/FEDER, UE)$]$; by the Walnuts and Healthy Aging study (http://www.clinicaltrials.gov; Grant NCT01634841) funded by the California Walnut Commission, Sacramento, California. BASE-II has been supported by the German Federal Ministry of Education and Research under grant numbers 16SV5537/\allowbreak 16SV5837/\allowbreak 16SV5538/\allowbreak 16SV5536K/\allowbreak 01UW0808/\allowbreak 01UW0706/\allowbreak 01GL1716A/\allowbreak 01GL1716B. Cam-CAN: Initial funding from the Biotechnology and Biological Sciences Research Council (BBSRC), followed by support from the Medical Research Council (MRC) Cognition \& Brain Sciences Unit (CBU).

Work on the Whitehall II Imaging Substudy was mainly funded by Lifelong Health and Wellbeing Programme Grant G1001354 from the UK Medical Research Council (“Predicting MRI Abnormalities with Longitudinal Data of the Whitehall II Substudy”) to Dr Ebmeier. The Wellcome Centre for Integrative Neuroimaging is supported by core funding from award 203139/Z/16/Z from the Wellcome Trust.

%

\begin{thebibliography}{82}
\providecommand{\natexlab}[1]{#1}
\providecommand{\url}[1]{\texttt{#1}}
\expandafter\ifx\csname urlstyle\endcsname\relax
  \providecommand{\doi}[1]{doi: #1}\else
  \providecommand{\doi}{doi: \begingroup \urlstyle{rm}\Url}\fi

\bibitem[Abellaneda-P{\'{e}}rez et~al.(2019)Abellaneda-P{\'{e}}rez,
  Vaqu{\'{e}}-Alc{\'{a}}zar, Vidal-Pi{\~{n}}eiro, Jannati, Solana,
  Bargall{\'{o}}, Santarnecchi, Pascual-Leone, and
  Bartr{\'{e}}s-Faz]{AbellanedaPrez2019}
Abellaneda-P{\'{e}}rez, Kilian; Vaqu{\'{e}}-Alc{\'{a}}zar, L{\'{\i}}dia;
  Vidal-Pi{\~{n}}eiro, D{\'{\i}}dac; Jannati, Ali; Solana, Elisabeth;
  Bargall{\'{o}}, N{\'{u}}ria; Santarnecchi, Emiliano; Pascual-Leone, Alvaro,
  and Bartr{\'{e}}s-Faz, David.
\newblock Age-related differences in default-mode network connectivity in
  response to intermittent theta-burst stimulation and its relationships with
  maintained cognition and brain integrity in healthy aging.
\newblock \emph{{NeuroImage}}, 188:\penalty0 794--806, March 2019.
\newblock \doi{10.1016/j.neuroimage.2018.11.036}.
\newblock URL \url{https://doi.org/10.1016/j.neuroimage.2018.11.036}.

\bibitem[Allen et~al.(2019)Allen, Poggiali, Whitaker, Marshall, and
  Kievit]{Allen2019}
Allen, Micah; Poggiali, Davide; Whitaker, Kirstie; Marshall, Tom~Rhys, and
  Kievit, Rogier~A.
\newblock Raincloud plots: a multi-platform tool for robust data visualization.
\newblock \emph{Wellcome Open Research}, 4:\penalty0 63, April 2019.
\newblock \doi{10.12688/wellcomeopenres.15191.1}.
\newblock URL \url{https://doi.org/10.12688/wellcomeopenres.15191.1}.

\bibitem[Bearden and Thompson(2017)]{Bearden2017}
Bearden, Carrie~E. and Thompson, Paul~M.
\newblock Emerging global initiatives in neurogenetics: The enhancing
  neuroimaging genetics through meta-analysis ({ENIGMA}) consortium.
\newblock \emph{Neuron}, 94\penalty0 (2):\penalty0 232--236, April 2017.
\newblock \doi{10.1016/j.neuron.2017.03.033}.
\newblock URL \url{https://doi.org/10.1016/j.neuron.2017.03.033}.

\bibitem[Becker(1994)]{Becker1994}
Becker, Betsy~Jane.
\newblock Combining significance levels.
\newblock In \emph{The handbook of research synthesis.}, pages 215--230.
  Russell Sage Foundation, New York, NY, US, 1994.
\newblock ISBN 0-87154-226-9 (Hardcover).

\bibitem[Berkey et~al.(1998)Berkey, Hoaglin, Antczak-Bouckoms, Mosteller, and
  Colditz]{Berkey1998}
Berkey, C.~S.; Hoaglin, D.~C.; Antczak-Bouckoms, A.; Mosteller, F., and
  Colditz, G.~A.
\newblock Meta-analysis of multiple outcomes by regression with random effects.
\newblock \emph{Statistics in Medicine}, 17\penalty0 (22):\penalty0 2537--2550,
  November 1998.
\newblock
  \doi{10.1002/(sici)1097-0258(19981130)17:22<2537::aid-sim953>3.0.co;2-c}.
\newblock URL
  \url{https://doi.org/10.1002/(sici)1097-0258(19981130)17:22<2537::aid-sim953>3.0.co;2-c}.

\bibitem[Bertram et~al.(2013)Bertram, B\"{o}ckenhoff, Demuth, D\"{u}zel,
  Eckardt, Li, Lindenberger, Pawelec, Siedler, Wagner, and
  Steinhagen-Thiessen]{Bertram2013}
Bertram, Lars; B\"{o}ckenhoff, Anke; Demuth, Ilja; D\"{u}zel, Sandra; Eckardt,
  Rahel; Li, Shu-Chen; Lindenberger, Ulman; Pawelec, Graham; Siedler, Thomas;
  Wagner, Gert~G, and Steinhagen-Thiessen, Elisabeth.
\newblock Cohort profile: The {Berlin} aging study {II}
  ({BASE}-{II}){\textdagger}.
\newblock \emph{International Journal of Epidemiology}, 43\penalty0
  (3):\penalty0 703--712, March 2013.
\newblock \doi{10.1093/ije/dyt018}.
\newblock URL \url{https://doi.org/10.1093/ije/dyt018}.

\bibitem[Birnbaum(1954)]{Birnbaum1954}
Birnbaum, Allan.
\newblock Combining independent tests of significance*.
\newblock \emph{Journal of the American Statistical Association}, 49\penalty0
  (267):\penalty0 559--574, 1954.
\newblock \doi{10.1080/01621459.1954.10483521}.
\newblock URL \url{https://doi.org/10.1080/01621459.1954.10483521}.

\bibitem[Boedhoe et~al.(2019)Boedhoe, Heymans, Schmaal, Abe, Alonso, Ameis,
  Anticevic, Arnold, Batistuzzo, Benedetti, Beucke, Bollettini, Bose, Brem,
  Calvo, Calvo, Cheng, Cho, Ciullo, Dallaspezia, Denys, Feusner, Fitzgerald,
  Fouche, Fridgeirsson, Gruner, Hanna, Hibar, Hoexter, Hu, Huyser, Jahanshad,
  James, Kathmann, Kaufmann, Koch, Kwon, Lazaro, Lochner, Marsh,
  Martínez-Zalacaín, Mataix-Cols, Menchón, Minuzzi, Morer, Nakamae, Nakao,
  Narayanaswamy, Nishida, Nurmi, O'Neill, Piacentini, Piras, Piras, Reddy,
  Reess, Sakai, Sato, Simpson, Soreni, Soriano-Mas, Spalletta, Stevens,
  Szeszko, Tolin, van Wingen, Venkatasubramanian, Walitza, Wang, Yun,
  Working-Group, Thompson, Stein, van~den Heuvel, and Twisk]{Boedhoe2019}
Boedhoe, Premika S.~W.; Heymans, Martijn~W.; Schmaal, Lianne; Abe, Yoshinari;
  Alonso, Pino; Ameis, Stephanie~H.; Anticevic, Alan; Arnold, Paul~D.;
  Batistuzzo, Marcelo~C.; Benedetti, Francesco; Beucke, Jan~C.; Bollettini,
  Irene; Bose, Anushree; Brem, Silvia; Calvo, Anna; Calvo, Rosa; Cheng, Yuqi;
  Cho, Kang Ik~K.; Ciullo, Valentina; Dallaspezia, Sara; Denys, Damiaan;
  Feusner, Jamie~D.; Fitzgerald, Kate~D.; Fouche, Jean-Paul; Fridgeirsson,
  Egill~A.; Gruner, Patricia; Hanna, Gregory~L.; Hibar, Derrek~P.; Hoexter,
  Marcelo~Q.; Hu, Hao; Huyser, Chaim; Jahanshad, Neda; James, Anthony;
  Kathmann, Norbert; Kaufmann, Christian; Koch, Kathrin; Kwon, Jun~Soo; Lazaro,
  Luisa; Lochner, Christine; Marsh, Rachel; Martínez-Zalacaín, Ignacio;
  Mataix-Cols, David; Menchón, José~M.; Minuzzi, Luciano; Morer, Astrid;
  Nakamae, Takashi; Nakao, Tomohiro; Narayanaswamy, Janardhanan~C.; Nishida,
  Seiji; Nurmi, Erika~L.; O'Neill, Joseph; Piacentini, John; Piras, Fabrizio;
  Piras, Federica; Reddy, Y.~C.~Janardhan; Reess, Tim~J.; Sakai, Yuki; Sato,
  Joao~R.; Simpson, H.~Blair; Soreni, Noam; Soriano-Mas, Carles; Spalletta,
  Gianfranco; Stevens, Michael~C.; Szeszko, Philip~R.; Tolin, David~F.; van
  Wingen, Guido~A.; Venkatasubramanian, Ganesan; Walitza, Susanne; Wang, Zhen;
  Yun, Je-Yeon; Working-Group, ENIGMA-OCD; Thompson, Paul~M.; Stein, Dan~J.;
  van~den Heuvel, Odile~A., and Twisk, Jos W.~R.
\newblock An empirical comparison of meta- and mega-analysis with data from the
  enigma obsessive-compulsive disorder working group.
\newblock \emph{Frontiers in Neuroinformatics}, 12:\penalty0 102, 2019.
\newblock ISSN 1662-5196.
\newblock \doi{10.3389/fninf.2018.00102}.
\newblock URL
  \url{https://www.frontiersin.org/article/10.3389/fninf.2018.00102}.

\bibitem[Borchers et~al.(1997)Borchers, Buckland, Priede, and
  Ahmadi]{Borchers1997}
Borchers, D~L; Buckland, S~T; Priede, I~G, and Ahmadi, S.
\newblock Improving the precision of the daily egg production method using
  generalized additive models.
\newblock \emph{Canadian Journal of Fisheries and Aquatic Sciences},
  54\penalty0 (12):\penalty0 2727--2742, 1997.
\newblock \doi{10.1139/f97-134}.
\newblock URL \url{https://doi.org/10.1139/f97-134}.

\bibitem[Brandmaier et~al.(2018)Brandmaier, von Oertzen, Ghisletta,
  Lindenberger, and Hertzog]{Brandmaier2018}
Brandmaier, Andreas~M.; von Oertzen, Timo; Ghisletta, Paolo; Lindenberger,
  Ulman, and Hertzog, Christopher.
\newblock Precision, reliability, and effect size of slope variance in latent
  growth curve models: Implications for statistical power analysis.
\newblock \emph{Frontiers in Psychology}, 9, April 2018.
\newblock \doi{10.3389/fpsyg.2018.00294}.
\newblock URL \url{https://doi.org/10.3389/fpsyg.2018.00294}.

\bibitem[Burke et~al.(2016)Burke, Ensor, and Riley]{Burke2016}
Burke, Danielle~L.; Ensor, Joie, and Riley, Richard~D.
\newblock Meta-analysis using individual participant data: one-stage and
  two-stage approaches, and why they may differ.
\newblock \emph{Statistics in Medicine}, 36\penalty0 (5):\penalty0 855--875,
  October 2016.
\newblock \doi{10.1002/sim.7141}.
\newblock URL \url{https://doi.org/10.1002/sim.7141}.

\bibitem[Buysse et~al.(1989)Buysse, Reynolds, Monk, Berman, and
  Kupfer]{Buysse1989}
Buysse, Daniel~J.; Reynolds, Charles~F.; Monk, Timothy~H.; Berman, Susan~R.,
  and Kupfer, David~J.
\newblock The {Pittsburgh} sleep quality index: A new instrument for
  psychiatric practice and research.
\newblock \emph{Psychiatry Research}, 28\penalty0 (2):\penalty0 193--213, May
  1989.
\newblock \doi{10.1016/0165-1781(89)90047-4}.
\newblock URL \url{https://doi.org/10.1016/0165-1781(89)90047-4}.

\bibitem[Cochran(1954)]{Cochran1954}
Cochran, William~G.
\newblock The combination of estimates from different experiments.
\newblock \emph{Biometrics}, 10\penalty0 (1):\penalty0 101, March 1954.
\newblock \doi{10.2307/3001666}.
\newblock URL \url{https://doi.org/10.2307/3001666}.

\bibitem[Crippa et~al.(2018)Crippa, Thomas, and Orsini]{Crippa2018}
Crippa, Alessio; Thomas, Ilias, and Orsini, Nicola.
\newblock A pointwise approach to dose-response meta-analysis of aggregated
  data.
\newblock \emph{International Journal of Statistics in Medical Research},
  7:\penalty0 25--32, 05 2018.
\newblock \doi{10.6000/1929-6029.2018.07.02.1}.

\bibitem[Dennis et~al.(2018)Dennis, Wilde, Newsome, Scheibel, Troyanskaya,
  Velez, Wade, Drennon, York, Bigler, Abildskov, Taylor, Jaramillo, Eapen,
  Belanger, Gupta, Morey, Haswell, Levin, Hinds, Walker, Thompson, and
  Tate]{Dennis2018}
Dennis, Emily~L.; Wilde, Elisabeth~A.; Newsome, Mary~R.; Scheibel, Randall~S.;
  Troyanskaya, Maya; Velez, Carmen; Wade, Benjamin~S.C.; Drennon, Ann~Marie;
  York, Gerald~E.; Bigler, Erin~D.; Abildskov, Tracy~J.; Taylor, Brian~A.;
  Jaramillo, Carlos~A.; Eapen, Blessen; Belanger, Heather; Gupta, Vikash;
  Morey, Rajendra; Haswell, Courtney; Levin, Harvey~S.; Hinds, Sidney~R.;
  Walker, William~C.; Thompson, Paul~M., and Tate, David~F.
\newblock {ENIGMA} military brain injury: A coordinated meta-analysis of
  diffusion {MRI} from multiple cohorts.
\newblock In \emph{2018 {IEEE} 15th International Symposium on Biomedical
  Imaging ({ISBI} 2018)}. {IEEE}, April 2018.
\newblock \doi{10.1109/isbi.2018.8363830}.
\newblock URL \url{https://doi.org/10.1109/isbi.2018.8363830}.

\bibitem[DerSimonian and Laird(1986)]{derSimonian1986}
DerSimonian, Rebecca and Laird, Nan.
\newblock Meta-analysis in clinical trials.
\newblock \emph{Controlled Clinical Trials}, 7\penalty0 (3):\penalty0 177 --
  188, 1986.
\newblock ISSN 0197-2456.
\newblock \doi{https://doi.org/10.1016/0197-2456(86)90046-2}.
\newblock URL
  \url{http://www.sciencedirect.com/science/article/pii/0197245686900462}.

\bibitem[Dewey(2019)]{Dewey2019}
Dewey, Michael.
\newblock \emph{{metap}: meta-analysis of significance values}, 2019.
\newblock R package version 1.2.

\bibitem[Edgington(1972)]{Edgington1972}
Edgington, Eugene~S.
\newblock An additive method for combining probability values from independent
  experiments.
\newblock \emph{The Journal of Psychology}, 80\penalty0 (2):\penalty0 351--363,
  1972.
\newblock \doi{10.1080/00223980.1972.9924813}.
\newblock URL \url{https://doi.org/10.1080/00223980.1972.9924813}.

\bibitem[Filippini et~al.(2014)Filippini, Zsoldos, Haapakoski, Sexton, Mahmood,
  Allan, Topiwala, Valkanova, Brunner, Shipley, Auerbach, Moeller,
  U{\u{g}}urbil, Xu, Yacoub, Andersson, Bijsterbosch, Clare, Griffanti, Hess,
  Jenkinson, Miller, Salimi-Khorshidi, Sotiropoulos, Voets, Smith, Geddes,
  Singh-Manoux, Mackay, Kivim\"{a}ki, and Ebmeier]{Filippini2014}
Filippini, Nicola; Zsoldos, Enik{\H{o}}; Haapakoski, Rita; Sexton, Claire~E;
  Mahmood, Abda; Allan, Charlotte~L; Topiwala, Anya; Valkanova, Vyara; Brunner,
  Eric~J; Shipley, Martin~J; Auerbach, Edward; Moeller, Steen; U{\u{g}}urbil,
  K{\^{a}}mil; Xu, Junqian; Yacoub, Essa; Andersson, Jesper; Bijsterbosch,
  Janine; Clare, Stuart; Griffanti, Ludovica; Hess, Aaron~T; Jenkinson, Mark;
  Miller, Karla~L; Salimi-Khorshidi, Gholamreza; Sotiropoulos, Stamatios~N;
  Voets, Natalie~L; Smith, Stephen~M; Geddes, John~R; Singh-Manoux, Archana;
  Mackay, Clare~E; Kivim\"{a}ki, Mika, and Ebmeier, Klaus~P.
\newblock Study protocol: the {Whitehall} {II} imaging sub-study.
\newblock \emph{{BMC} Psychiatry}, 14\penalty0 (1), May 2014.
\newblock \doi{10.1186/1471-244x-14-159}.
\newblock URL \url{https://doi.org/10.1186/1471-244x-14-159}.

\bibitem[Fisher(1925)]{Fisher1925}
Fisher, RA.
\newblock \emph{Statistical methods for research workers}.
\newblock Oliver and Boyd, Edinburgh, 1925.

\bibitem[Fjell et~al.(2010)Fjell, Walhovd, Westlye, {\O}stby, Tamnes, Jernigan,
  Gamst, and Dale]{Fjell2010}
Fjell, Anders.~M.; Walhovd, Kristine~B.; Westlye, Lars~T.; {\O}stby, Ylva;
  Tamnes, Christian~K.; Jernigan, Terry~L.; Gamst, Anthony, and Dale, Anders~M.
\newblock When does brain aging accelerate? dangers of quadratic fits in
  cross-sectional studies.
\newblock \emph{NeuroImage}, 50\penalty0 (4):\penalty0 1376 -- 1383, 2010.
\newblock ISSN 1053-8119.
\newblock \doi{https://doi.org/10.1016/j.neuroimage.2010.01.061}.
\newblock URL
  \url{http://www.sciencedirect.com/science/article/pii/S1053811910000832}.

\bibitem[Fjell et~al.(2019)Fjell, S{\o}rensen, Amlien, Bartr{\'{e}}s-Faz, Bros,
  Buchmann, Demuth, Drevon, D\"{u}zel, Ebmeier, Idland, Kietzmann, Kievit,
  K\"{u}hn, Lindenberger, Mowinckel, Nyberg, Price, Sexton,
  Sol{\'{e}}-Padull{\'{e}}s, Pudas, Sederevicius, Suri, Wagner, Watne,
  Westerhausen, Zsoldos, and Walhovd]{Fjell2019}
Fjell, Anders~M; S{\o}rensen, {\O}ystein; Amlien, Inge~K; Bartr{\'{e}}s-Faz,
  David; Bros, Didac~Maci{\'{a}}; Buchmann, Nikolaus; Demuth, Ilja; Drevon,
  Christian~A; D\"{u}zel, Sandra; Ebmeier, Klaus~P; Idland, Ane-Victoria;
  Kietzmann, Tim~C; Kievit, Rogier; K\"{u}hn, Simone; Lindenberger, Ulman;
  Mowinckel, Athanasia~M; Nyberg, Lars; Price, Darren; Sexton, Claire~E;
  Sol{\'{e}}-Padull{\'{e}}s, Cristina; Pudas, Sara; Sederevicius, Donatas;
  Suri, Sana; Wagner, Gerd; Watne, Leiv~Otto; Westerhausen, Ren{\'{e}};
  Zsoldos, Enik{\H{o}}, and Walhovd, Kristine~B.
\newblock Self-reported sleep relates to hippocampal atrophy across the adult
  lifespan {\textendash} results from the lifebrain consortium.
\newblock \emph{Sleep}, November 2019.
\newblock \doi{10.1093/sleep/zsz280}.
\newblock URL \url{https://doi.org/10.1093/sleep/zsz280}.

\bibitem[Fjell et~al.(2017)Fjell, Idland, Sala-Llonch, Watne, Borza,
  Br{\ae}khus, Lona, Zetterberg, Blennow, Wyller, and Walhovd]{Fjell2017}
Fjell, Anders~Martin; Idland, Ane-Victoria; Sala-Llonch, Roser; Watne,
  Leiv~Otto; Borza, Tom; Br{\ae}khus, Anne; Lona, Tarjei; Zetterberg, Henrik;
  Blennow, Kaj; Wyller, Torgeir~Bruun, and Walhovd, Kristine~Beate.
\newblock Neuroinflammation and tau interact with amyloid in predicting sleep
  problems in aging independently of atrophy.
\newblock \emph{Cerebral Cortex}, 28\penalty0 (8):\penalty0 2775--2785, June
  2017.
\newblock \doi{10.1093/cercor/bhx157}.
\newblock URL \url{https://doi.org/10.1093/cercor/bhx157}.

\bibitem[Foster and Deardorff(2017)]{FosterMSLS2017}
Foster, Erin~D. and Deardorff, Ariel.
\newblock Open science framework ({OSF}).
\newblock \emph{Journal of the Medical Library Association}, 105\penalty0 (2),
  April 2017.
\newblock \doi{10.5195/jmla.2017.88}.
\newblock URL \url{https://doi.org/10.5195/jmla.2017.88}.

\bibitem[Gasparrini et~al.(2012)Gasparrini, Armstrong, and
  Kenward]{Gasparrini2012}
Gasparrini, A.; Armstrong, B., and Kenward, M.~G.
\newblock Multivariate meta-analysis for non-linear and other multi-parameter
  associations.
\newblock \emph{Statistics in Medicine}, 31\penalty0 (29):\penalty0 3821--3839,
  2012.
\newblock \doi{10.1002/sim.5471}.
\newblock URL \url{https://onlinelibrary.wiley.com/doi/abs/10.1002/sim.5471}.

\bibitem[Gasparrini and Armstrong(2010)]{Gasparrini2010}
Gasparrini, Antonio and Armstrong, Ben.
\newblock Time series analysis on the health effects of temperature:
  Advancements and limitations.
\newblock \emph{Environmental Research}, 110\penalty0 (6):\penalty0 633 -- 638,
  2010.
\newblock ISSN 0013-9351.
\newblock \doi{https://doi.org/10.1016/j.envres.2010.06.005}.
\newblock URL
  \url{http://www.sciencedirect.com/science/article/pii/S0013935110000976}.

\bibitem[Gerstorf et~al.(2016)Gerstorf, Bertram, Lindenberger, Pawelec, Demuth,
  Steinhagen-Thiessen, and Wagner]{Gerstorf2016}
Gerstorf, Denis; Bertram, Lars; Lindenberger, Ulman; Pawelec, Graham; Demuth,
  Ilja; Steinhagen-Thiessen, Elisabeth, and Wagner, Gert~G.
\newblock Editorial.
\newblock \emph{Gerontology}, 62\penalty0 (3):\penalty0 311--315, 2016.
\newblock \doi{10.1159/000441495}.
\newblock URL \url{https://doi.org/10.1159/000441495}.

\bibitem[Harrell(2019)]{Harrel2019}
Harrell, Frank~E.
\newblock \emph{Hmisc: Harrell Miscellaneous}, 2019.
\newblock URL \url{https://CRAN.R-project.org/package=Hmisc}.
\newblock R package version 4.3-0.

\bibitem[Hastie and Tibshirani(1986)]{Hastie1986}
Hastie, Trevor and Tibshirani, Robert.
\newblock Generalized additive models.
\newblock \emph{Statist. Sci.}, 1\penalty0 (3):\penalty0 297--310, 08 1986.
\newblock \doi{10.1214/ss/1177013604}.
\newblock URL \url{https://doi.org/10.1214/ss/1177013604}.

\bibitem[Hastie and Tibshirani(1993)]{Hastie1993}
Hastie, Trevor and Tibshirani, Robert.
\newblock Varying-coefficient models.
\newblock \emph{Journal of the Royal Statistical Society: Series B
  (Methodological)}, 55\penalty0 (4):\penalty0 757--779, September 1993.
\newblock \doi{10.1111/j.2517-6161.1993.tb01939.x}.
\newblock URL \url{https://doi.org/10.1111/j.2517-6161.1993.tb01939.x}.

\bibitem[Hedges et~al.(2010)Hedges, Tipton, and Johnson]{Hedges2010}
Hedges, Larry~V.; Tipton, Elizabeth, and Johnson, Matthew~C.
\newblock {Robust variance estimation in meta‐regression with dependent
  effect size estimates}.
\newblock \emph{Research Synthesis Methods}, 1\penalty0 (1):\penalty0 39--65, 1
  2010.
\newblock ISSN 1759-2887.
\newblock \doi{10.1002/jrsm.5}.
\newblock URL \url{https://doi.org/10.1002/jrsm.5}.

\bibitem[Hedges and Olkin(1985)]{Hedges1985}
Hedges, L.V. and Olkin, I.
\newblock \emph{Statistical methods for meta-analysis}.
\newblock Academic Press, Orland, FL, 1985.

\bibitem[Hofer and Piccinin(2009)]{Hofer2009}
Hofer, Scott~M. and Piccinin, Andrea~M.
\newblock Integrative data analysis through coordination of measurement and
  analysis protocol across independent longitudinal studies.
\newblock \emph{Psychological Methods}, 14\penalty0 (2):\penalty0 150--164,
  2009.
\newblock \doi{10.1037/a0015566}.
\newblock URL \url{https://doi.org/10.1037/a0015566}.

\bibitem[Kievit et~al.(2018)Kievit, Brandmaier, Ziegler, van Harmelen,
  de~Mooij, Moutoussis, Goodyer, Bullmore, Jones, Fonagy, Lindenberger, and
  Dolan]{Kievit2018}
Kievit, Rogier~A.; Brandmaier, Andreas~M.; Ziegler, Gabriel; van Harmelen,
  Anne-Laura; de~Mooij, Susanne~M.M.; Moutoussis, Michael; Goodyer, Ian~M.;
  Bullmore, Ed; Jones, Peter~B.; Fonagy, Peter; Lindenberger, Ulman, and Dolan,
  Raymond~J.
\newblock Developmental cognitive neuroscience using latent change score
  models: A tutorial and applications.
\newblock \emph{Developmental Cognitive Neuroscience}, 33:\penalty0 99--117,
  October 2018.
\newblock \doi{10.1016/j.dcn.2017.11.007}.
\newblock URL \url{https://doi.org/10.1016/j.dcn.2017.11.007}.

\bibitem[Kontopantelis(2018)]{Kontopantelis2018}
Kontopantelis, Evangelos.
\newblock A comparison of one-stage vs two-stage individual patient data
  meta-analysis methods: A simulation study.
\newblock \emph{Research Synthesis Methods}, June 2018.
\newblock \doi{10.1002/jrsm.1303}.
\newblock URL \url{https://doi.org/10.1002/jrsm.1303}.

\bibitem[Laird and Ware(1982)]{Laird1982}
Laird, Nan~M. and Ware, James~H.
\newblock Random-effects models for longitudinal data.
\newblock \emph{Biometrics}, 38\penalty0 (4):\penalty0 963--974, 1982.
\newblock ISSN 0006341X, 15410420.
\newblock URL \url{http://www.jstor.org/stable/2529876}.

\bibitem[Little(1993)]{Little1993}
Little, Roderick J.~A.
\newblock Statistical analysis of masked data.
\newblock \emph{Journal of Official Statistics}, 9\penalty0 (2):\penalty0 407,
  06 1993.
\newblock URL
  \url{https://search.proquest.com/docview/1266808565?accountid=14699}.

\bibitem[Loughin(2004)]{Loughin2004}
Loughin, Thomas~M.
\newblock A systematic comparison of methods for combining p-values from
  independent tests.
\newblock \emph{Computational Statistics \& Data Analysis}, 47\penalty0
  (3):\penalty0 467 -- 485, 2004.
\newblock ISSN 0167-9473.
\newblock \doi{https://doi.org/10.1016/j.csda.2003.11.020}.
\newblock URL
  \url{http://www.sciencedirect.com/science/article/pii/S0167947303002950}.

\bibitem[Marra and Wood(2012)]{Marra2012}
Marra, Giamperio and Wood, Simon~N.
\newblock Coverage properties of confidence intervals for generalized additive
  model components.
\newblock \emph{Scandinavian Journal of Statistics}, 39\penalty0 (1):\penalty0
  53--74, 2012.
\newblock \doi{10.1111/j.1467-9469.2011.00760.x}.
\newblock URL
  \url{https://onlinelibrary.wiley.com/doi/abs/10.1111/j.1467-9469.2011.00760.x}.

\bibitem[McArdle and Horn(1985)]{McArdle1985}
McArdle, JJ and Horn, JL.
\newblock Mega analyses of the wais: Structural and dynamic models of adult
  intellectual ability.
\newblock \emph{National institute of aging grant report, University of
  Viginia}, 1985.

\bibitem[McCullagh and Nelder(1989)]{McCullagh1989}
McCullagh, P. and Nelder, J.~A.
\newblock \emph{Generalized Linear Models}.
\newblock Chapman \& Hall / CRC, London, 1989.

\bibitem[Murdoch et~al.(2008)Murdoch, Tsai, and Adcock]{Murdoch2008}
Murdoch, Duncan~J; Tsai, Yu-Ling, and Adcock, James.
\newblock P-values are random variables.
\newblock \emph{The American Statistician}, 62\penalty0 (3):\penalty0 242--245,
  August 2008.
\newblock \doi{10.1198/000313008x332421}.
\newblock URL \url{https://doi.org/10.1198/000313008x332421}.

\bibitem[Murphy(2012)]{Murphy2012}
Murphy, Kevin~P.
\newblock \emph{Machine Learning: A Probabilistic Perspective}.
\newblock The MIT Press, 2012.
\newblock ISBN 0262018020, 9780262018029.

\bibitem[Nilsson et~al.(1997)Nilsson, B\"{a}ckman, Erngrund, Nyberg, Adolfsson,
  Bucht, Karlsson, Widing, and Winblad]{Nilsson1997}
Nilsson, Lars-G\"{o}ran; B\"{a}ckman, Lars; Erngrund, Karin; Nyberg, Lars;
  Adolfsson, Rolf; Bucht, G\"{o}sta; Karlsson, Stig; Widing, Maud, and Winblad,
  Bengt.
\newblock The {B}etula prospective cohort study: Memory, health, and aging.
\newblock \emph{Aging, Neuropsychology, and Cognition}, 4\penalty0
  (1):\penalty0 1--32, March 1997.
\newblock \doi{10.1080/13825589708256633}.
\newblock URL \url{https://doi.org/10.1080/13825589708256633}.

\bibitem[Nowok et~al.(2016)Nowok, Raab, and Dibben]{Nowok2016}
Nowok, Beata; Raab, Gillian~M., and Dibben, Chris.
\newblock synthpop: Bespoke creation of synthetic data in {R}.
\newblock \emph{Journal of Statistical Software}, 74\penalty0 (11), 2016.
\newblock \doi{10.18637/jss.v074.i11}.
\newblock URL \url{https://doi.org/10.18637/jss.v074.i11}.

\bibitem[Nychka(1988)]{Nychka1988}
Nychka, Douglas.
\newblock Bayesian confidence intervals for smoothing splines.
\newblock \emph{Journal of the American Statistical Association}, 83\penalty0
  (404):\penalty0 1134--1143, December 1988.
\newblock \doi{10.1080/01621459.1988.10478711}.
\newblock URL \url{https://doi.org/10.1080/01621459.1988.10478711}.

\bibitem[Pedersen et~al.(2019)Pedersen, Miller, Simpson, and
  Ross]{Pedersen2019}
Pedersen, Eric~J.; Miller, David~L.; Simpson, Gavin~L., and Ross, Noam.
\newblock Hierarchical generalized additive models in ecology: an introduction
  with mgcv.
\newblock \emph{{PeerJ}}, 7:\penalty0 e6876, May 2019.
\newblock \doi{10.7717/peerj.6876}.
\newblock URL \url{https://doi.org/10.7717/peerj.6876}.

\bibitem[Piers(2018)]{Piers2018}
Piers, Ryan~J.
\newblock Structural brain volume differences between cognitively intact
  {ApoE}4 carriers and non-carriers across the lifespan.
\newblock \emph{Neural Regeneration Research}, 13\penalty0 (8):\penalty0 1309,
  2018.
\newblock \doi{10.4103/1673-5374.235408}.
\newblock URL \url{https://doi.org/10.4103/1673-5374.235408}.

\bibitem[Pinheiro et~al.(2019)Pinheiro, Bates, DebRoy, Sarkar, and {R Core
  Team}]{Pinheiro2019}
Pinheiro, Jose; Bates, Douglas; DebRoy, Saikat; Sarkar, Deepayan, and {R Core
  Team}, .
\newblock \emph{{nlme}: Linear and Nonlinear Mixed Effects Models}, 2019.
\newblock URL \url{https://CRAN.R-project.org/package=nlme}.
\newblock R package version 3.1-143.

\bibitem[Polanin et~al.(2017)Polanin, Hennessy, and Tanner-Smith]{Polanin2017}
Polanin, Joshua~R.; Hennessy, Emily~A., and Tanner-Smith, Emily~E.
\newblock A review of meta-analysis packages in {R}.
\newblock \emph{Journal of Educational and Behavioral Statistics}, 42\penalty0
  (2):\penalty0 206--242, 2017.
\newblock \doi{10.3102/1076998616674315}.
\newblock URL \url{https://doi.org/10.3102/1076998616674315}.

\bibitem[{R Core Team}(2019)]{Rcore}
{R Core Team}, .
\newblock \emph{R: A Language and Environment for Statistical Computing}.
\newblock R Foundation for Statistical Computing, Vienna, Austria, 2019.
\newblock URL \url{https://www.R-project.org/}.

\bibitem[Rajaram et~al.(2017)Rajaram, Valls-Pedret, Cof{\'{a}}n, Sabat{\'{e}},
  Serra-Mir, P{\'{e}}rez-Heras, Arechiga, Casaroli-Marano, Alforja, Sala-Vila,
  Dom{\'{e}}nech, Roth, Freitas-Simoes, Calvo, L{\'{o}}pez-Illamola, Haddad,
  Bitok, Kazzi, Huey, Fan, and Ros]{Rajaram2017}
Rajaram, Sujatha; Valls-Pedret, Cinta; Cof{\'{a}}n, Montserrat; Sabat{\'{e}},
  Joan; Serra-Mir, Merc{\`{e}}; P{\'{e}}rez-Heras, Ana~M.; Arechiga, Adam;
  Casaroli-Marano, Ricardo~P.; Alforja, Socorro; Sala-Vila, Aleix;
  Dom{\'{e}}nech, M{\'{o}}nica; Roth, Irene; Freitas-Simoes, Tania~M.; Calvo,
  Carlos; L{\'{o}}pez-Illamola, Anna; Haddad, Ella; Bitok, Edward; Kazzi,
  Natalie; Huey, Lynnley; Fan, Joseph, and Ros, Emilio.
\newblock The walnuts and healthy aging study ({WAHA}): Protocol for a
  nutritional intervention trial with walnuts on brain aging.
\newblock \emph{Frontiers in Aging Neuroscience}, 8, January 2017.
\newblock \doi{10.3389/fnagi.2016.00333}.
\newblock URL \url{https://doi.org/10.3389/fnagi.2016.00333}.

\bibitem[Riley et~al.(2010)Riley, Lambert, and Abo-Zaid]{Riley2010}
Riley, Richard~D; Lambert, Paul~C, and Abo-Zaid, Ghada.
\newblock Meta-analysis of individual participant data: rationale, conduct, and
  reporting.
\newblock \emph{BMJ}, 340, 2010.
\newblock ISSN 0959-8138.
\newblock \doi{10.1136/bmj.c221}.
\newblock URL \url{https://www.bmj.com/content/340/bmj.c221}.

\bibitem[Rosenthal(1978)]{Rosenthal1978}
Rosenthal, Robert.
\newblock Combining results of independent studies.
\newblock \emph{Psychological Bulletin}, 85\penalty0 (1):\penalty0 185--193,
  1978.
\newblock \doi{10.1037/0033-2909.85.1.185}.
\newblock URL \url{https://doi.org/10.1037/0033-2909.85.1.185}.

\bibitem[Rubin(1993)]{Rubin1993}
Rubin, Donald~B.
\newblock Discussion statistical disclosure limitation.
\newblock \emph{Journal of Official Statistics}, 9\penalty0 (2):\penalty0 461,
  06 1993.
\newblock URL
  \url{https://search-proquest-com.ezproxy.uio.no/docview/1266818482?accountid=14699}.

\bibitem[Salimi-Khorshidi et~al.(2011)Salimi-Khorshidi, Nichols, Smith, and
  Woolrich]{SalimiKhorshidi2011}
Salimi-Khorshidi, G.; Nichols, T.~E.; Smith, S.~M., and Woolrich, M.~W.
\newblock Using {Gaussian}-process regression for meta-analytic neuroimaging
  inference based on sparse observations.
\newblock \emph{{IEEE} Transactions on Medical Imaging}, 30\penalty0
  (7):\penalty0 1401--1416, July 2011.
\newblock \doi{10.1109/tmi.2011.2122341}.
\newblock URL \url{https://doi.org/10.1109/tmi.2011.2122341}.

\bibitem[Sauerbrei and Royston(2011)]{Sauerbrei2011}
Sauerbrei, Willi and Royston, Patrick.
\newblock A new strategy for meta-analysis of continuous covariates in
  observational studies.
\newblock \emph{Statistics in Medicine}, 30\penalty0 (28):\penalty0 3341--3360,
  2011.
\newblock \doi{10.1002/sim.4333}.
\newblock URL \url{https://onlinelibrary.wiley.com/doi/abs/10.1002/sim.4333}.

\bibitem[Schwartz and Zanobetti(2000)]{Schwarz2000}
Schwartz, Joel and Zanobetti, Antonella.
\newblock Using meta-smoothing to estimate dose-response trends across multiple
  studies, with application to air pollution and daily death.
\newblock \emph{Epidemiology}, 11\penalty0 (6):\penalty0 666--672, 2000.
\newblock ISSN 10443983.
\newblock URL \url{http://www.jstor.org/stable/3703820}.

\bibitem[Servén and Brummitt(2018)]{Serven2018}
Servén, Daniel and Brummitt, Charlie.
\newblock pygam: Generalized additive models in python, March 2018.
\newblock URL \url{https://doi.org/10.5281/zenodo.1208723}.

\bibitem[S{\o}rensen et~al.(2020)S{\o}rensen, Brandmaier, and
  Mowinckel]{Sorensen2020}
S{\o}rensen, {\O}ystein; Brandmaier, Andreas~M., and Mowinckel, Athanasia~Mo.
\newblock \emph{metagam: Meta-analysis of generalized additive models}, 2020.
\newblock URL \url{https://CRAN.R-project.org/package=metagam}.
\newblock R package version 0.1.2.

\bibitem[Stouffer et~al.(1949)Stouffer, Suchman, DeVinney, Star, and
  Williams]{Stouffer1949}
Stouffer, SA; Suchman, EA; DeVinney, LC; Star, SA, and Williams, RMJ.
\newblock \emph{The American soldier, vol 1: Adjustment during army life}.
\newblock Princeton University Press, Princeton, 1949.

\bibitem[Sung et~al.(2014)Sung, Schwander, Arnett, Kardia, Rankinen, Bouchard,
  Boerwinkle, Hunt, and Rao]{Sung2014}
Sung, Yun~Ju; Schwander, Karen; Arnett, Donna~K.; Kardia, Sharon~L.R.;
  Rankinen, Tuomo; Bouchard, Claude; Boerwinkle, Eric; Hunt, Steven~C., and
  Rao, Dabeeru~C.
\newblock An empirical comparison of meta-analysis and mega-analysis of
  individual participant data for identifying gene-environment interactions.
\newblock \emph{Genetic Epidemiology}, 38\penalty0 (4):\penalty0 369--378,
  April 2014.
\newblock \doi{10.1002/gepi.21800}.
\newblock URL \url{https://doi.org/10.1002/gepi.21800}.

\bibitem[Sutton and Higgins(2008)]{Sutton2008}
Sutton, Alexander~J. and Higgins, Julian P.~T.
\newblock Recent developments in meta-analysis.
\newblock \emph{Statistics in Medicine}, 27\penalty0 (5):\penalty0 625--650,
  2008.
\newblock \doi{10.1002/sim.2934}.
\newblock URL \url{https://onlinelibrary.wiley.com/doi/abs/10.1002/sim.2934}.

\bibitem[Taylor et~al.(2017)Taylor, Williams, Cusack, Auer, Shafto, Dixon,
  Tyler, Cam-CAN, and Henson]{Taylor2017}
Taylor, Jason~R.; Williams, Nitin; Cusack, Rhodri; Auer, Tibor; Shafto,
  Meredith~A.; Dixon, Marie; Tyler, Lorraine~K.; Cam-CAN, , and Henson,
  Richard~N.
\newblock The cambridge centre for ageing and neuroscience (cam-{CAN}) data
  repository: Structural and functional {MRI}, {MEG}, and cognitive data from a
  cross-sectional adult lifespan sample.
\newblock \emph{{NeuroImage}}, 144:\penalty0 262--269, January 2017.
\newblock \doi{10.1016/j.neuroimage.2015.09.018}.
\newblock URL \url{https://doi.org/10.1016/j.neuroimage.2015.09.018}.

\bibitem[Thompson et~al.(2014)Thompson, , Stein, Medland, Hibar, Vasquez,
  Renteria, Toro, Jahanshad, Schumann, Franke, Wright, Martin, Agartz, Alda,
  Alhusaini, Almasy, Almeida, Alpert, Andreasen, Andreassen, Apostolova, Appel,
  Armstrong, Aribisala, Bastin, Bauer, Bearden, Bergmann, Binder, Blangero,
  Bockholt, B{\o}en, Bois, Boomsma, Booth, Bowman, Bralten, Brouwer, Brunner,
  Brohawn, Buckner, Buitelaar, Bulayeva, Bustillo, Calhoun, Cannon, Cantor,
  Carless, Caseras, Cavalleri, Chakravarty, Chang, Ching, Christoforou, Cichon,
  Clark, Conrod, Coppola, Crespo-Facorro, Curran, Czisch, Deary, de~Geus, den
  Braber, Delvecchio, Depondt, de~Haan, de~Zubicaray, Dima, Dimitrova,
  Djurovic, Dong, Donohoe, Duggirala, Dyer, Ehrlich, Ekman, Elvs{\aa}shagen,
  Emsell, Erk, Espeseth, Fagerness, Fears, Fedko, Fern{\'{a}}ndez, Fisher,
  Foroud, Fox, Francks, Frangou, Frey, Frodl, Frouin, Garavan, Giddaluru,
  Glahn, Godlewska, Goldstein, Gollub, Grabe, Grimm, Gruber, Guadalupe, Gur,
  Gur, G\"{o}ring, Hagenaars, Hajek, Hall, Hall, Hardy, Hartman, Hass, Hatton,
  Haukvik, Hegenscheid, Heinz, Hickie, Ho, Hoehn, Hoekstra, Hollinshead,
  Holmes, Homuth, Hoogman, Hong, Hosten, Hottenga, Pol, Hwang, Jack, Jenkinson,
  Johnston, J\"{o}nsson, Kahn, Kasperaviciute, Kelly, Kim, Kochunov, Koenders,
  Kr\"{a}mer, Kwok, Lagopoulos, Laje, Landen, Landman, Lauriello, Lawrie, Lee,
  Hellard, Lema{\^{\i}}tre, Leonardo, shan Li, Liberg, Liewald, Liu, Lopez,
  Loth, Lourdusamy, Luciano, Macciardi, Machielsen, MacQueen, Malt, Mandl,
  Manoach, Martinot, Matarin, Mather, Mattheisen, Mattingsdal,
  Meyer-Lindenberg, McDonald, McIntosh, McMahon, McMahon, Meisenzahl, Melle,
  Milaneschi, Mohnke, Montgomery, Morris, Moses, Mueller, Maniega,
  M\"{u}hleisen, M\"{u}ller-Myhsok, Mwangi, Nauck, Nho, Nichols, Nilsson,
  Nugent, Nyberg, Olvera, Oosterlaan, Ophoff, Pandolfo,
  Papalampropoulou-Tsiridou, Papmeyer, Paus, Pausova, Pearlson, Penninx,
  Peterson, Pfennig, Phillips, Pike, Poline, Potkin, P\"{u}tz, Ramasamy,
  Rasmussen, Rietschel, Rijpkema, Risacher, Roffman, Roiz-Santia{\~{n}}ez,
  Romanczuk-Seiferth, Rose, Royle, Rujescu, Ryten, Sachdev, Salami,
  Satterthwaite, Savitz, Saykin, Scanlon, Schmaal, Schnack, Schork, Schulz,
  Sch\"{u}r, Seidman, Shen, Shoemaker, Simmons, Sisodiya, Smith, Smoller,
  Soares, Sponheim, Sprooten, Starr, Steen, Strakowski, Strike, Sussmann,
  S\"{a}mann, Teumer, Toga, Tordesillas-Gutierrez, Trabzuni, Trost, Turner, den
  Heuvel, van~der Wee, van Eijk, van Erp, van Haren, van~`t Ent, van Tol,
  Hern{\'{a}}ndez, Veltman, Versace, V\"{o}lzke, Walker, Walter, Wang, Wardlaw,
  Weale, Weiner, Wen, Westlye, Whalley, Whelan, White, Winkler, Wittfeld,
  Woldehawariat, Wolf, Zilles, Zwiers, Thalamuthu, Schofield, Freimer,
  Lawrence, and Drevets]{Thompson2014}
Thompson, Paul~M.; ; Stein, Jason~L.; Medland, Sarah~E.; Hibar, Derrek~P.;
  Vasquez, Alejandro~Arias; Renteria, Miguel~E.; Toro, Roberto; Jahanshad,
  Neda; Schumann, Gunter; Franke, Barbara; Wright, Margaret~J.; Martin,
  Nicholas~G.; Agartz, Ingrid; Alda, Martin; Alhusaini, Saud; Almasy, Laura;
  Almeida, Jorge; Alpert, Kathryn; Andreasen, Nancy~C.; Andreassen, Ole~A.;
  Apostolova, Liana~G.; Appel, Katja; Armstrong, Nicola~J.; Aribisala,
  Benjamin; Bastin, Mark~E.; Bauer, Michael; Bearden, Carrie~E.; Bergmann,
  {\O}rjan; Binder, Elisabeth~B.; Blangero, John; Bockholt, Henry~J.; B{\o}en,
  Erlend; Bois, Catherine; Boomsma, Dorret~I.; Booth, Tom; Bowman, Ian~J.;
  Bralten, Janita; Brouwer, Rachel~M.; Brunner, Han~G.; Brohawn, David~G.;
  Buckner, Randy~L.; Buitelaar, Jan; Bulayeva, Kazima; Bustillo, Juan~R.;
  Calhoun, Vince~D.; Cannon, Dara~M.; Cantor, Rita~M.; Carless, Melanie~A.;
  Caseras, Xavier; Cavalleri, Gianpiero~L.; Chakravarty, M.~Mallar; Chang,
  Kiki~D.; Ching, Christopher R.~K.; Christoforou, Andrea; Cichon, Sven; Clark,
  Vincent~P.; Conrod, Patricia; Coppola, Giovanni; Crespo-Facorro, Benedicto;
  Curran, Joanne~E.; Czisch, Michael; Deary, Ian~J.; de~Geus, Eco J.~C.; den
  Braber, Anouk; Delvecchio, Giuseppe; Depondt, Chantal; de~Haan, Lieuwe;
  de~Zubicaray, Greig~I.; Dima, Danai; Dimitrova, Rali; Djurovic, Srdjan; Dong,
  Hongwei; Donohoe, Gary; Duggirala, Ravindranath; Dyer, Thomas~D.; Ehrlich,
  Stefan; Ekman, Carl~Johan; Elvs{\aa}shagen, Torbj{\o}rn; Emsell, Louise; Erk,
  Susanne; Espeseth, Thomas; Fagerness, Jesen; Fears, Scott; Fedko, Iryna;
  Fern{\'{a}}ndez, Guill{\'{e}}n; Fisher, Simon~E.; Foroud, Tatiana; Fox,
  Peter~T.; Francks, Clyde; Frangou, Sophia; Frey, Eva~Maria; Frodl, Thomas;
  Frouin, Vincent; Garavan, Hugh; Giddaluru, Sudheer; Glahn, David~C.;
  Godlewska, Beata; Goldstein, Rita~Z.; Gollub, Randy~L.; Grabe, Hans~J.;
  Grimm, Oliver; Gruber, Oliver; Guadalupe, Tulio; Gur, Raquel~E.; Gur,
  Ruben~C.; G\"{o}ring, Harald H.~H.; Hagenaars, Saskia; Hajek, Tomas; Hall,
  Geoffrey~B.; Hall, Jeremy; Hardy, John; Hartman, Catharina~A.; Hass, Johanna;
  Hatton, Sean~N.; Haukvik, Unn~K.; Hegenscheid, Katrin; Heinz, Andreas;
  Hickie, Ian~B.; Ho, Beng-Choon; Hoehn, David; Hoekstra, Pieter~J.;
  Hollinshead, Marisa; Holmes, Avram~J.; Homuth, Georg; Hoogman, Martine; Hong,
  L.~Elliot; Hosten, Norbert; Hottenga, Jouke-Jan; Pol, Hilleke E.~Hulshoff;
  Hwang, Kristy~S.; Jack, Clifford~R.; Jenkinson, Mark; Johnston, Caroline;
  J\"{o}nsson, Erik~G.; Kahn, Ren{\'{e}}~S.; Kasperaviciute, Dalia; Kelly,
  Sinead; Kim, Sungeun; Kochunov, Peter; Koenders, Laura; Kr\"{a}mer, Bernd;
  Kwok, John B.~J.; Lagopoulos, Jim; Laje, Gonzalo; Landen, Mikael; Landman,
  Bennett~A.; Lauriello, John; Lawrie, Stephen~M.; Lee, Phil~H.; Hellard,
  Stephanie~Le; Lema{\^{\i}}tre, Herve; Leonardo, Cassandra~D.; shan Li,
  Chiang; Liberg, Benny; Liewald, David~C.; Liu, Xinmin; Lopez, Lorna~M.; Loth,
  Eva; Lourdusamy, Anbarasu; Luciano, Michelle; Macciardi, Fabio; Machielsen,
  Marise W.~J.; MacQueen, Glenda~M.; Malt, Ulrik~F.; Mandl, Ren{\'{e}};
  Manoach, Dara~S.; Martinot, Jean-Luc; Matarin, Mar; Mather, Karen~A.;
  Mattheisen, Manuel; Mattingsdal, Morten; Meyer-Lindenberg, Andreas; McDonald,
  Colm; McIntosh, Andrew~M.; McMahon, Francis~J.; McMahon, Katie~L.;
  Meisenzahl, Eva; Melle, Ingrid; Milaneschi, Yuri; Mohnke, Sebastian;
  Montgomery, Grant~W.; Morris, Derek~W.; Moses, Eric~K.; Mueller, Bryon~A.;
  Maniega, Susana~Mu{\~{n}}oz; M\"{u}hleisen, Thomas~W.; M\"{u}ller-Myhsok,
  Bertram; Mwangi, Benson; Nauck, Matthias; Nho, Kwangsik; Nichols, Thomas~E.;
  Nilsson, Lars-G\"{o}ran; Nugent, Allison~C.; Nyberg, Lars; Olvera, Rene~L.;
  Oosterlaan, Jaap; Ophoff, Roel~A.; Pandolfo, Massimo;
  Papalampropoulou-Tsiridou, Melina; Papmeyer, Martina; Paus, Tomas; Pausova,
  Zdenka; Pearlson, Godfrey~D.; Penninx, Brenda~W.; Peterson, Charles~P.;
  Pfennig, Andrea; Phillips, Mary; Pike, G.~Bruce; Poline, Jean-Baptiste;
  Potkin, Steven~G.; P\"{u}tz, Benno; Ramasamy, Adaikalavan; Rasmussen, Jerod;
  Rietschel, Marcella; Rijpkema, Mark; Risacher, Shannon~L.; Roffman,
  Joshua~L.; Roiz-Santia{\~{n}}ez, Roberto; Romanczuk-Seiferth, Nina; Rose,
  Emma~J.; Royle, Natalie~A.; Rujescu, Dan; Ryten, Mina; Sachdev, Perminder~S.;
  Salami, Alireza; Satterthwaite, Theodore~D.; Savitz, Jonathan; Saykin,
  Andrew~J.; Scanlon, Cathy; Schmaal, Lianne; Schnack, Hugo~G.; Schork,
  Andrew~J.; Schulz, S.~Charles; Sch\"{u}r, Remmelt; Seidman, Larry; Shen, Li;
  Shoemaker, Jody~M.; Simmons, Andrew; Sisodiya, Sanjay~M.; Smith, Colin;
  Smoller, Jordan~W.; Soares, Jair~C.; Sponheim, Scott~R.; Sprooten, Emma;
  Starr, John~M.; Steen, Vidar~M.; Strakowski, Stephen; Strike, Lachlan;
  Sussmann, Jessika; S\"{a}mann, Philipp~G.; Teumer, Alexander; Toga,
  Arthur~W.; Tordesillas-Gutierrez, Diana; Trabzuni, Daniah; Trost, Sarah;
  Turner, Jessica; den Heuvel, Martijn~Van; van~der Wee, Nic~J.; van Eijk,
  Kristel; van Erp, Theo G.~M.; van Haren, Neeltje E.~M.; van~`t Ent, Dennis;
  van Tol, Marie-Jose; Hern{\'{a}}ndez, Maria C.~Vald{\'{e}}s; Veltman,
  Dick~J.; Versace, Amelia; V\"{o}lzke, Henry; Walker, Robert; Walter, Henrik;
  Wang, Lei; Wardlaw, Joanna~M.; Weale, Michael~E.; Weiner, Michael~W.; Wen,
  Wei; Westlye, Lars~T.; Whalley, Heather~C.; Whelan, Christopher~D.; White,
  Tonya; Winkler, Anderson~M.; Wittfeld, Katharina; Woldehawariat, Girma; Wolf,
  Christiane; Zilles, David; Zwiers, Marcel~P.; Thalamuthu, Anbupalam;
  Schofield, Peter~R.; Freimer, Nelson~B.; Lawrence, Natalia~S., and Drevets,
  Wayne.
\newblock The {ENIGMA} consortium: large-scale collaborative analyses of
  neuroimaging and genetic data.
\newblock \emph{Brain Imaging and Behavior}, January 2014.
\newblock \doi{10.1007/s11682-013-9269-5}.
\newblock URL \url{https://doi.org/10.1007/s11682-013-9269-5}.

\bibitem[Thompson et~al.(2017)Thompson, Andreassen, Arias-Vasquez, Bearden,
  Boedhoe, Brouwer, Buckner, Buitelaar, Bulayeva, Cannon, Cohen, Conrod, Dale,
  Deary, Dennis, de~Reus, Desrivieres, Dima, Donohoe, Fisher, Fouche, Francks,
  Frangou, Franke, Ganjgahi, Garavan, Glahn, Grabe, Guadalupe, Gutman,
  Hashimoto, Hibar, Holland, Hoogman, Pol, Hosten, Jahanshad, Kelly, Kochunov,
  Kremen, Lee, Mackey, Martin, Mazoyer, McDonald, Medland, Morey, Nichols,
  Paus, Pausova, Schmaal, Schumann, Shen, Sisodiya, Smit, Smoller, Stein,
  Stein, Toro, Turner, van~den Heuvel, van~den Heuvel, van Erp, van Rooij,
  Veltman, Walter, Wang, Wardlaw, Whelan, Wright, and Ye]{Thompson2017}
Thompson, Paul~M.; Andreassen, Ole~A.; Arias-Vasquez, Alejandro; Bearden,
  Carrie~E.; Boedhoe, Premika~S.; Brouwer, Rachel~M.; Buckner, Randy~L.;
  Buitelaar, Jan~K.; Bulayeva, Kazima~B.; Cannon, Dara~M.; Cohen, Ronald~A.;
  Conrod, Patricia~J.; Dale, Anders~M.; Deary, Ian~J.; Dennis, Emily~L.;
  de~Reus, Marcel~A.; Desrivieres, Sylvane; Dima, Danai; Donohoe, Gary; Fisher,
  Simon~E.; Fouche, Jean-Paul; Francks, Clyde; Frangou, Sophia; Franke,
  Barbara; Ganjgahi, Habib; Garavan, Hugh; Glahn, David~C.; Grabe, Hans~J.;
  Guadalupe, Tulio; Gutman, Boris~A.; Hashimoto, Ryota; Hibar, Derrek~P.;
  Holland, Dominic; Hoogman, Martine; Pol, Hilleke E.~Hulshoff; Hosten,
  Norbert; Jahanshad, Neda; Kelly, Sinead; Kochunov, Peter; Kremen, William~S.;
  Lee, Phil~H.; Mackey, Scott; Martin, Nicholas~G.; Mazoyer, Bernard; McDonald,
  Colm; Medland, Sarah~E.; Morey, Rajendra~A.; Nichols, Thomas~E.; Paus, Tomas;
  Pausova, Zdenka; Schmaal, Lianne; Schumann, Gunter; Shen, Li; Sisodiya,
  Sanjay~M.; Smit, Dirk~J.A.; Smoller, Jordan~W.; Stein, Dan~J.; Stein,
  Jason~L.; Toro, Roberto; Turner, Jessica~A.; van~den Heuvel, Martijn~P.;
  van~den Heuvel, Odile~L.; van Erp, Theo~G.M.; van Rooij, Daan; Veltman,
  Dick~J.; Walter, Henrik; Wang, Yalin; Wardlaw, Joanna~M.; Whelan,
  Christopher~D.; Wright, Margaret~J., and Ye, Jieping.
\newblock {ENIGMA} and the individual: Predicting factors that affect the brain
  in 35 countries worldwide.
\newblock \emph{{NeuroImage}}, 145:\penalty0 389--408, January 2017.
\newblock \doi{10.1016/j.neuroimage.2015.11.057}.
\newblock URL \url{https://doi.org/10.1016/j.neuroimage.2015.11.057}.

\bibitem[Tippet(1931)]{Tippet1931}
Tippet, LHC.
\newblock \emph{The Methods of Statistics}.
\newblock Williams and Norgate, London, 1931.

\bibitem[van Erp et~al.(2018)van Erp, Walton, Hibar, Schmaal, Jiang, Glahn,
  Pearlson, Yao, Fukunaga, Hashimoto, Okada, Yamamori, Bustillo, Clark, Agartz,
  Mueller, Cahn, de~Zwarte, Pol, Kahn, Ophoff, van Haren, Andreassen, Dale,
  Doan, Gurholt, Hartberg, Haukvik, J{\o}rgensen, Lagerberg, Melle, Westlye,
  Gruber, Kraemer, Richter, Zilles, Calhoun, Crespo-Facorro,
  Roiz-Santia{\~{n}}ez, Tordesillas-Guti{\'{e}}rrez, Loughland, Carr, Catts,
  Cropley, Fullerton, Green, Henskens, Jablensky, Lenroot, Mowry, Michie,
  Pantelis, Quid{\'{e}}, Schall, Scott, Cairns, Seal, Tooney, Rasser, Cooper,
  Weickert, Weickert, Morris, Hong, Kochunov, Beard, Gur, Gur, Satterthwaite,
  Wolf, Belger, Brown, Ford, Macciardi, Mathalon, O'Leary, Potkin, Preda,
  Voyvodic, Lim, McEwen, Yang, Tan, Tan, Wang, Fan, Chen, Xiang, Tang, Guo,
  Wan, Wei, Bockholt, Ehrlich, Wolthusen, King, Shoemaker, Sponheim, Haan,
  Koenders, Machielsen, van Amelsvoort, Veltman, Assogna, Banaj, de~Rossi,
  Iorio, Piras, Spalletta, McKenna, Pomarol-Clotet, Salvador, Corvin, Donohoe,
  Kelly, Whelan, Dickie, Rotenberg, Voineskos, Ciufolini, Radua, Dazzan,
  Murray, Marques, Simmons, Borgwardt, Egloff, Harrisberger,
  Riecher-R\"{o}ssler, Smieskova, Alpert, Wang, J\"{o}nsson, Koops, Sommer,
  Bertolino, Bonvino, Giorgio, Neilson, Mayer, Stephen, Kwon, Yun, Cannon,
  McDonald, Lebedeva, Tomyshev, Akhadov, Kaleda, Fatouros-Bergman, Flyckt,
  Busatto, Rosa, Serpa, Zanetti, Hoschl, Skoch, Spaniel, Tomecek, Hagenaars,
  McIntosh, Whalley, Lawrie, Kn\"{o}chel, Oertel-Kn\"{o}chel, St\"{a}blein,
  Howells, Stein, Temmingh, Uhlmann, Lopez-Jaramillo, Dima, McMahon, Faskowitz,
  Gutman, Jahanshad, Thompson, Turner, Farde, Flyckt, Engberg, Erhardt,
  Fatouros-Bergman, Cervenka, Schwieler, Piehl, Agartz, Collste, Victorsson,
  Malmqvist, Hedberg, and Orhan]{vanErp2018}
van Erp, Theo~G.M.; Walton, Esther; Hibar, Derrek~P.; Schmaal, Lianne; Jiang,
  Wenhao; Glahn, David~C.; Pearlson, Godfrey~D.; Yao, Nailin; Fukunaga, Masaki;
  Hashimoto, Ryota; Okada, Naohiro; Yamamori, Hidenaga; Bustillo, Juan~R.;
  Clark, Vincent~P.; Agartz, Ingrid; Mueller, Bryon~A.; Cahn, Wiepke;
  de~Zwarte, Sonja~M.C.; Pol, Hilleke E.~Hulshoff; Kahn, Ren{\'{e}}~S.; Ophoff,
  Roel~A.; van Haren, Neeltje~E.M.; Andreassen, Ole~A.; Dale, Anders~M.; Doan,
  Nhat~Trung; Gurholt, Tiril~P.; Hartberg, Cecilie~B.; Haukvik, Unn~K.;
  J{\o}rgensen, Kjetil~N.; Lagerberg, Trine~V.; Melle, Ingrid; Westlye,
  Lars~T.; Gruber, Oliver; Kraemer, Bernd; Richter, Anja; Zilles, David;
  Calhoun, Vince~D.; Crespo-Facorro, Benedicto; Roiz-Santia{\~{n}}ez, Roberto;
  Tordesillas-Guti{\'{e}}rrez, Diana; Loughland, Carmel; Carr, Vaughan~J.;
  Catts, Stanley; Cropley, Vanessa~L.; Fullerton, Janice~M.; Green, Melissa~J.;
  Henskens, Frans~A.; Jablensky, Assen; Lenroot, Rhoshel~K.; Mowry, Bryan~J.;
  Michie, Patricia~T.; Pantelis, Christos; Quid{\'{e}}, Yann; Schall, Ulrich;
  Scott, Rodney~J.; Cairns, Murray~J.; Seal, Marc; Tooney, Paul~A.; Rasser,
  Paul~E.; Cooper, Gavin; Weickert, Cynthia~Shannon; Weickert, Thomas~W.;
  Morris, Derek~W.; Hong, Elliot; Kochunov, Peter; Beard, Lauren~M.; Gur,
  Raquel~E.; Gur, Ruben~C.; Satterthwaite, Theodore~D.; Wolf, Daniel~H.;
  Belger, Aysenil; Brown, Gregory~G.; Ford, Judith~M.; Macciardi, Fabio;
  Mathalon, Daniel~H.; O'Leary, Daniel~S.; Potkin, Steven~G.; Preda, Adrian;
  Voyvodic, James; Lim, Kelvin~O.; McEwen, Sarah; Yang, Fude; Tan, Yunlong;
  Tan, Shuping; Wang, Zhiren; Fan, Fengmei; Chen, Jingxu; Xiang, Hong; Tang,
  Shiyou; Guo, Hua; Wan, Ping; Wei, Dong; Bockholt, Henry~J.; Ehrlich, Stefan;
  Wolthusen, Rick~P.F.; King, Margaret~D.; Shoemaker, Jody~M.; Sponheim,
  Scott~R.; Haan, Lieuwe~De; Koenders, Laura; Machielsen, Marise~W.; van
  Amelsvoort, Therese; Veltman, Dick~J.; Assogna, Francesca; Banaj, Nerisa;
  de~Rossi, Pietro; Iorio, Mariangela; Piras, Fabrizio; Spalletta, Gianfranco;
  McKenna, Peter~J.; Pomarol-Clotet, Edith; Salvador, Raymond; Corvin, Aiden;
  Donohoe, Gary; Kelly, Sinead; Whelan, Christopher~D.; Dickie, Erin~W.;
  Rotenberg, David; Voineskos, Aristotle~N.; Ciufolini, Simone; Radua, Joaquim;
  Dazzan, Paola; Murray, Robin; Marques, Tiago~Reis; Simmons, Andrew;
  Borgwardt, Stefan; Egloff, Laura; Harrisberger, Fabienne;
  Riecher-R\"{o}ssler, Anita; Smieskova, Renata; Alpert, Kathryn~I.; Wang, Lei;
  J\"{o}nsson, Erik~G.; Koops, Sanne; Sommer, Iris~E.C.; Bertolino, Alessandro;
  Bonvino, Aurora; Giorgio, Annabella~Di; Neilson, Emma; Mayer, Andrew~R.;
  Stephen, Julia~M.; Kwon, Jun~Soo; Yun, Je-Yeon; Cannon, Dara~M.; McDonald,
  Colm; Lebedeva, Irina; Tomyshev, Alexander~S.; Akhadov, Tolibjohn; Kaleda,
  Vasily; Fatouros-Bergman, Helena; Flyckt, Lena; Busatto, Geraldo~F.; Rosa,
  Pedro~G.P.; Serpa, Mauricio~H.; Zanetti, Marcus~V.; Hoschl, Cyril; Skoch,
  Antonin; Spaniel, Filip; Tomecek, David; Hagenaars, Saskia~P.; McIntosh,
  Andrew~M.; Whalley, Heather~C.; Lawrie, Stephen~M.; Kn\"{o}chel, Christian;
  Oertel-Kn\"{o}chel, Viola; St\"{a}blein, Michael; Howells, Fleur~M.; Stein,
  Dan~J.; Temmingh, Henk~S.; Uhlmann, Anne; Lopez-Jaramillo, Carlos; Dima,
  Danai; McMahon, Agnes; Faskowitz, Joshua~I.; Gutman, Boris~A.; Jahanshad,
  Neda; Thompson, Paul~M.; Turner, Jessica~A.; Farde, Lars; Flyckt, Lena;
  Engberg, G\"{o}ran; Erhardt, Sophie; Fatouros-Bergman, Helena; Cervenka,
  Simon; Schwieler, Lilly; Piehl, Fredrik; Agartz, Ingrid; Collste, Karin;
  Victorsson, Pauliina; Malmqvist, Anna; Hedberg, Mikael, and Orhan, Funda.
\newblock Cortical brain abnormalities in 4474 individuals with schizophrenia
  and 5098 control subjects via the enhancing neuro imaging genetics through
  meta analysis ({ENIGMA}) consortium.
\newblock \emph{Biological Psychiatry}, 84\penalty0 (9):\penalty0 644--654,
  November 2018.
\newblock \doi{10.1016/j.biopsych.2018.04.023}.
\newblock URL \url{https://doi.org/10.1016/j.biopsych.2018.04.023}.

\bibitem[Veroniki et~al.(2016)Veroniki, Jackson, Viechtbauer, Bender, Bowden,
  Knapp, Kuss, Higgins, Langan, and Salanti]{Veroniki2016}
Veroniki, Areti~Angeliki; Jackson, Dan; Viechtbauer, Wolfgang; Bender, Ralf;
  Bowden, Jack; Knapp, Guido; Kuss, Oliver; Higgins, Julian~PT; Langan, Dean,
  and Salanti, Georgia.
\newblock Methods to estimate the between-study variance and its uncertainty in
  meta-analysis.
\newblock \emph{Research Synthesis Methods}, 7\penalty0 (1):\penalty0 55--79,
  2016.
\newblock \doi{10.1002/jrsm.1164}.
\newblock URL \url{https://onlinelibrary.wiley.com/doi/abs/10.1002/jrsm.1164}.

\bibitem[Vidal-Pi{\~{n}}eiro et~al.(2014)Vidal-Pi{\~{n}}eiro, Martin-Trias,
  Arenaza-Urquijo, Sala-Llonch, Clemente, Mena-S{\'{a}}nchez, Bargall{\'{o}},
  Falc{\'{o}}n, Pascual-Leone, and Bartr{\'{e}}s-Faz]{VidalPineiro2014}
Vidal-Pi{\~{n}}eiro, D{\'{\i}}dac; Martin-Trias, Pablo; Arenaza-Urquijo,
  Eider~M.; Sala-Llonch, Roser; Clemente, Imma~C.; Mena-S{\'{a}}nchez, Isaias;
  Bargall{\'{o}}, N{\'{u}}ria; Falc{\'{o}}n, Carles; Pascual-Leone,
  {\'{A}}lvaro, and Bartr{\'{e}}s-Faz, David.
\newblock Task-dependent activity and connectivity predict episodic memory
  network-based responses to brain stimulation in healthy aging.
\newblock \emph{Brain Stimulation}, 7\penalty0 (2):\penalty0 287--296, March
  2014.
\newblock \doi{10.1016/j.brs.2013.12.016}.
\newblock URL \url{https://doi.org/10.1016/j.brs.2013.12.016}.

\bibitem[Viechtbauer(2005)]{Viechtbauer2005}
Viechtbauer, Wolfgang.
\newblock Bias and efficiency of meta-analytic variance estimators in the
  random-effects model.
\newblock \emph{Journal of Educational and Behavioral Statistics}, 30\penalty0
  (3):\penalty0 261--293, September 2005.
\newblock \doi{10.3102/10769986030003261}.
\newblock URL \url{https://doi.org/10.3102/10769986030003261}.

\bibitem[Viechtbauer(2010)]{Viechtbauer2010}
Viechtbauer, Wolfgang.
\newblock Conducting meta-analyses in {R} with the metafor package.
\newblock \emph{Journal of Statistical Software, Articles}, 36\penalty0
  (3):\penalty0 1--48, 2010.
\newblock ISSN 1548-7660.
\newblock \doi{10.18637/jss.v036.i03}.
\newblock URL \url{https://www.jstatsoft.org/v036/i03}.

\bibitem[Viechtbauer et~al.(2015)Viechtbauer, L{\'{o}}pez-L{\'{o}}pez,
  S{\'{a}}nchez-Meca, and Mar{\'{\i}}n-Mart{\'{\i}}nez]{Viechtbauer2015}
Viechtbauer, Wolfgang; L{\'{o}}pez-L{\'{o}}pez, Jos{\'{e}}~Antonio;
  S{\'{a}}nchez-Meca, Julio, and Mar{\'{\i}}n-Mart{\'{\i}}nez, Fulgencio.
\newblock A comparison of procedures to test for moderators in mixed-effects
  meta-regression models.
\newblock \emph{Psychological Methods}, 20\penalty0 (3):\penalty0 360--374,
  2015.
\newblock \doi{10.1037/met0000023}.
\newblock URL \url{https://doi.org/10.1037/met0000023}.

\bibitem[Walhovd et~al.(2018)Walhovd, Fjell, Westerhausen, Nyberg, Ebmeier,
  Lindenberger, Bartres-Faz, Baare, Siebner, Henson, and et~al.]{Walhovd2018}
Walhovd, K.B.; Fjell, A.M.; Westerhausen, R.; Nyberg, L.; Ebmeier, K.P.;
  Lindenberger, U.; Bartres-Faz, D.; Baare, W.F.C.; Siebner, H.R.; Henson, R.,
  and et~al., .
\newblock Healthy minds 0–100 years: Optimising the use of {European} brain
  imaging cohorts (“{Lifebrain}”).
\newblock \emph{European Psychiatry}, 47:\penalty0 76–77, 2018.
\newblock \doi{10.1016/j.eurpsy.2017.10.005}.

\bibitem[Walhovd et~al.(2016)Walhovd, Krogsrud, Amlien, Bartsch, Bj{\o}rnerud,
  Due-T{\o}nnessen, Grydeland, Hagler, H{\aa}berg, Kremen, Ferschmann, Nyberg,
  Panizzon, Rohani, Skranes, Storsve, S{\o}lsnes, Tamnes, Thompson, Reuter,
  Dale, and Fjell]{Walhovd2016}
Walhovd, Kristine~B.; Krogsrud, Stine~K.; Amlien, Inge~K.; Bartsch, Hauke;
  Bj{\o}rnerud, Atle; Due-T{\o}nnessen, Paulina; Grydeland, H{\aa}kon; Hagler,
  Donald~J.; H{\aa}berg, Asta~K.; Kremen, William~S.; Ferschmann, Lia; Nyberg,
  Lars; Panizzon, Matthew~S.; Rohani, Darius~A.; Skranes, Jon; Storsve,
  Andreas~B.; S{\o}lsnes, Anne~Elisabeth; Tamnes, Christian~K.; Thompson,
  Wesley~K.; Reuter, Chase; Dale, Anders~M., and Fjell, Anders~M.
\newblock Neurodevelopmental origins of lifespan changes in brain and
  cognition.
\newblock \emph{Proceedings of the National Academy of Sciences}, 113\penalty0
  (33):\penalty0 9357--9362, July 2016.
\newblock \doi{10.1073/pnas.1524259113}.
\newblock URL \url{https://doi.org/10.1073/pnas.1524259113}.

\bibitem[Walhovd et~al.(2019)Walhovd, Fjell, S{\o}rensen, Mowinckel, Reinbold,
  Idland, Watne, Franke, Dobricic, Kilpert, Bertram, and Wang]{Walhovd2019}
Walhovd, Kristine~B; Fjell, Anders~M.; S{\o}rensen, {\O}ystein; Mowinckel,
  Athanasia~Monica; Reinbold, C{\'e}line~Sonja; Idland, Ane-Victoria; Watne,
  Leiv~Otto; Franke, Andre; Dobricic, Valerijia; Kilpert, Fabian; Bertram,
  Lars, and Wang, Yunpeng.
\newblock Genetic risk for {A}lzheimer{\textquoteright}s disease predicts
  hippocampal volume through the lifespan.
\newblock \emph{bioRxiv}, 2019.
\newblock \doi{10.1101/711689}.
\newblock URL \url{https://www.biorxiv.org/content/early/2019/07/23/711689}.

\bibitem[Wickham(2016)]{Wickham2016}
Wickham, Hadley.
\newblock \emph{ggplot2: Elegant Graphics for Data Analysis}.
\newblock Springer-Verlag New York, 2016.
\newblock ISBN 978-3-319-24277-4.
\newblock URL \url{https://ggplot2.tidyverse.org}.

\bibitem[Wilkinson(1951)]{Wilkinson1951}
Wilkinson, Bryan.
\newblock A statistical consideration in psychological research.
\newblock \emph{Psychological Bulletin}, 48\penalty0 (2):\penalty0 156--158,
  1951.
\newblock \doi{10.1037/h0059111}.
\newblock URL \url{https://doi.org/10.1037/h0059111}.

\bibitem[Wood and Scheipl(2017)]{Wood2017gamm4}
Wood, Simon and Scheipl, Fabian.
\newblock \emph{gamm4: Generalized Additive Mixed Models using 'mgcv' and
  'lme4'}, 2017.
\newblock URL \url{https://CRAN.R-project.org/package=gamm4}.
\newblock R package version 0.2-5.

\bibitem[Wood(2012)]{Wood2012}
Wood, Simon~N.
\newblock {On p-values for smooth components of an extended generalized
  additive model}.
\newblock \emph{Biometrika}, 100\penalty0 (1):\penalty0 221--228, 10 2012.
\newblock ISSN 0006-3444.
\newblock \doi{10.1093/biomet/ass048}.
\newblock URL \url{https://doi.org/10.1093/biomet/ass048}.

\bibitem[Wood(2017)]{Wood2017}
Wood, S.N.
\newblock \emph{Generalized Additive Models: An Introduction with R}.
\newblock Chapman and Hall/CRC, 2 edition, 2017.

\bibitem[Zaykin(2011)]{Zaykin2011}
Zaykin, D.~V.
\newblock Optimally weighted z-test is a powerful method for combining
  probabilities in meta-analysis.
\newblock \emph{Journal of Evolutionary Biology}, 24\penalty0 (8):\penalty0
  1836--1841, 2011.
\newblock \doi{10.1111/j.1420-9101.2011.02297.x}.
\newblock URL
  \url{https://onlinelibrary.wiley.com/doi/abs/10.1111/j.1420-9101.2011.02297.x}.

\end{thebibliography}

\appendix

\section{Identifiability Constraints on Smooth Terms}
\label{sec:smooth_constraints}

The smooth terms in the GAM \eqref{eq:gam_generic} are only uniquely determined up to some additive constant. In order to compute the model fit, constraints have to be imposed on the smooth terms, effectively fixing $f_{s}(0)$ to some constant value. The default in the R package \verb!mgcv! is to let each smooth term $f_{s}(\mathcal{X}_{s})$ sum to zero over the observed data $\mathcal{X}_{s}$. This means requiring that the smooth term estimated from data in each cohort satisfy
\begin{equation}
\label{eq:sum_to_zero}
\sum_{\mathbf{x} \in \mathcal{X}_{s,m}} {f}_{s,m}\left(\mathbf{x}\right) = 0, ~ m=1,\dots,M,
\end{equation}
where we let $\mathcal{X}_{s,m}$ denote the actual values of $\mathcal{X}_{s}$ in cohort $m$. Using this approach the smooth term in each cohort has been constrained to sum to zero over its own data, and hence the terms are not directly comparable without correcting for this difference in offset. This is particularly important when the values of $\mathcal{X}_{s,m}$ cover different ranges across cohorts, as in Figure \ref{fig:cohort_dist}.

One solution is to note that the smooth plus its intercept are comparable across cohorts, since the difference between the constraints is captured by the intercept term. To be precise, assume a GAM with a single smooth term $f_{1}$ is  fit to data in cohorts $m_{1}$ and $m_{2}$, where the smooth term is constrained according to the data in cohort $m_{1}$, i.e.,
\begin{equation*}
\sum_{\mathbf{x} \in \mathcal{X}_{s,m_{1}}} {f}_{s,m}\left(\mathbf{x}\right) = 0, ~ m = m_{1}, m_{2}.
\end{equation*}
This yields estimates $\tilde{\beta}_{0,m} + \tilde{f}_{s,m}$ for $m=m_{1}, m_{2}$, and the terms $\tilde{f}_{s,m_{1}}$ and $\tilde{f}_{s,m_{2}}$ would be directly comparable. Instead constraining ${f}_{s,m_{2}}$ over its own data would lead to a shift $\Delta \tilde{\beta}_{0,m_{2}}$ in the intercept estimated in cohort $m_{2}$, i.e.,
\begin{equation*}
\sum_{\mathbf{x} \in \mathcal{X}_{s,m_{1}}} {f}_{s,m_{2}}(\mathbf{x})= \Delta \tilde{\beta}_{0,m_{2}}+
\sum_{\mathbf{x} \in \mathcal{X}_{s,m_{2}}} {f}_{s,m_{2}}(\mathbf{x}) =0.
\end{equation*}
The estimated intercept in cohort $m_{2}$ would now be $\hat{\beta}_{0,m_{2}} = \tilde{\beta}_{0,m_{2}} + \Delta \tilde{\beta}_{0,m_{2}}$, where $\Delta \tilde{\beta}_{0,m_{2}}$ takes into account the difference between the sum-to-zero constraint in cohort $m_{1}$ and in cohort $m_{2}$. This argument generalizes to any number of cohorts and smooth terms. Hence, sum-to-zero constraints of the form \eqref{eq:sum_to_zero} for each smooth can be imposed independently in each cohort fit, as long as the estimated intercept $\beta_{0}$ is added to each smooth term before combining. This implies replacing $\hat{f}_{s,m}$ with $\hat{\beta}_{0,m} + \hat{f}_{s,m}$ in equation \eqref{eq:meta_est}. An important point for interpretation is that when using this option, the meta-analytic estimate of $\beta_{0} + f_{s}$ incorporates both differences between estimated intercepts and differences between estimated smooth terms across cohorts.

Another way to resolve this issue is by imposing a constraint for each smooth term, specifying a point at which it should be exactly zero \citep[Ch. 5.4.1]{Wood2017}. If the same point constraints have been applied when fitting the GAM to the data from each cohort, the smooth terms are all on the same scale and can be combined meta-analytically as described in Section \ref{sec:methods}. This approach hence replaces \eqref{eq:sum_to_zero} by
\begin{equation}
\label{eq:point_constraint}
f_{s, m}\left(\mathcal{X}_{s}^{pc}\right) = 0, ~ m=1,\dots,M,
\end{equation}
for some point $\mathcal{X}_{s}^{pc}$ which is identical across cohorts. An advantage of this approach is that it does not require the intercept to be included in the meta-analysis; hence the meta-analytic estimate $\hat{f}_{s}$ contains only the smooth term. On the other hand, point constraint may lead to wider confidence bands for the smooth terms \citep[Ch. 5.4.1]{Wood2017}. Also, this approach requires that point constraints are specified as part of the model to be fit to the data from each cohort. Note that the confidence interval for a smooth term subject to point constraint \eqref{eq:point_constraint} does not need to have zero width at the constraint point $\mathcal{X}_{s}^{pc}$. The methods for constructing confidence intervals developed by \citet{Marra2012} based on the work by \citet{Nychka1988}, take into account the uncertainty about the overall intercept as well as the uncertainty about the smooth term, and these typically yield better coverage properties than confidence intervals which only model the uncertainty of the smooth term.

\end{document}